\documentclass{article} 
\usepackage[table]{xcolor}
\usepackage{iclr2026_conference,times}

\usepackage{amsmath,amsfonts,bm}

\def\eqref#1{equation~\ref{#1}}

\def\1{\bm{1}}

\DeclareMathAlphabet{\mathsfit}{\encodingdefault}{\sfdefault}{m}{sl}
\SetMathAlphabet{\mathsfit}{bold}{\encodingdefault}{\sfdefault}{bx}{n}

\usepackage{url}
\usepackage{graphicx}
\usepackage{booktabs}
\usepackage{multirow}
\usepackage{enumitem}
\usepackage{array}
\usepackage{float}
\usepackage{caption}
\usepackage{subcaption}
\usepackage{wrapfig}
\usepackage{amssymb}
\usepackage{xcolor}
\definecolor{Highlight}{rgb}{0.21,0.49,0.74}
\definecolor{red}{HTML}{cc1100}
\definecolor{mygray}{gray}{0.95}
\usepackage[pagebackref,breaklinks,colorlinks=true,linkcolor=red,citecolor=Highlight]{hyperref}  
\usepackage[T1]{fontenc}
\usepackage[raster,skins]{tcolorbox}
\setlist[itemize]{leftmargin=16pt, itemsep=0.15ex, topsep=0.15ex, parsep=0pt, partopsep=0pt}

\iclrfinalcopy

\title{MGM-Omni: Scaling Omni LLMs to Personalized Long-Horizon Speech}

\author{
\begin{minipage}[t]{\textwidth}
\centering
Chengyao Wang$^1$\thanks{equal contribution} \quad
Zhisheng Zhong$^1$\footnotemark[1] \quad
Bohao Peng$^1$\footnotemark[1] \quad
Senqiao Yang$^1$ \quad
Yuqi Liu$^1$ \\[1mm]
\centering
Haokun Gui$^2$ \quad
Bin Xia$^1$ \quad
Jingyao Li$^1$ \quad
Bei Yu$^1$ \quad
Jiaya Jia$^{23}$ \\[3mm]
\centering
$^1$CUHK \quad $^2$HKUST \quad $^3$SmartMore \\[3mm]
\includegraphics[height=1.0em]{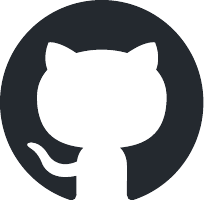}\hspace{0.5em}{\small \url{https://github.com/dvlab-research/MGM-Omni}} \\[0.5mm]
\includegraphics[height=1.0em]{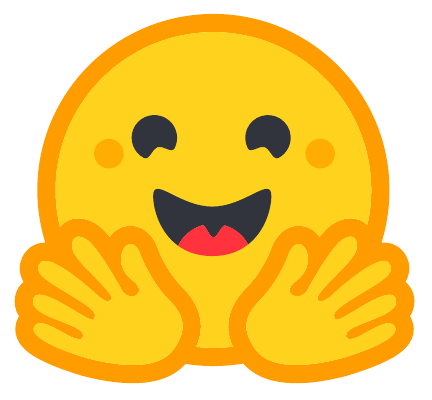}\hspace{0.5em}{\small \url{https://huggingface.co/spaces/wcy1122/MGM-Omni}}
\end{minipage}
}

\begin{document}

\maketitle

\begin{figure}[h]
\begin{center}
\includegraphics[width=1.0\textwidth]{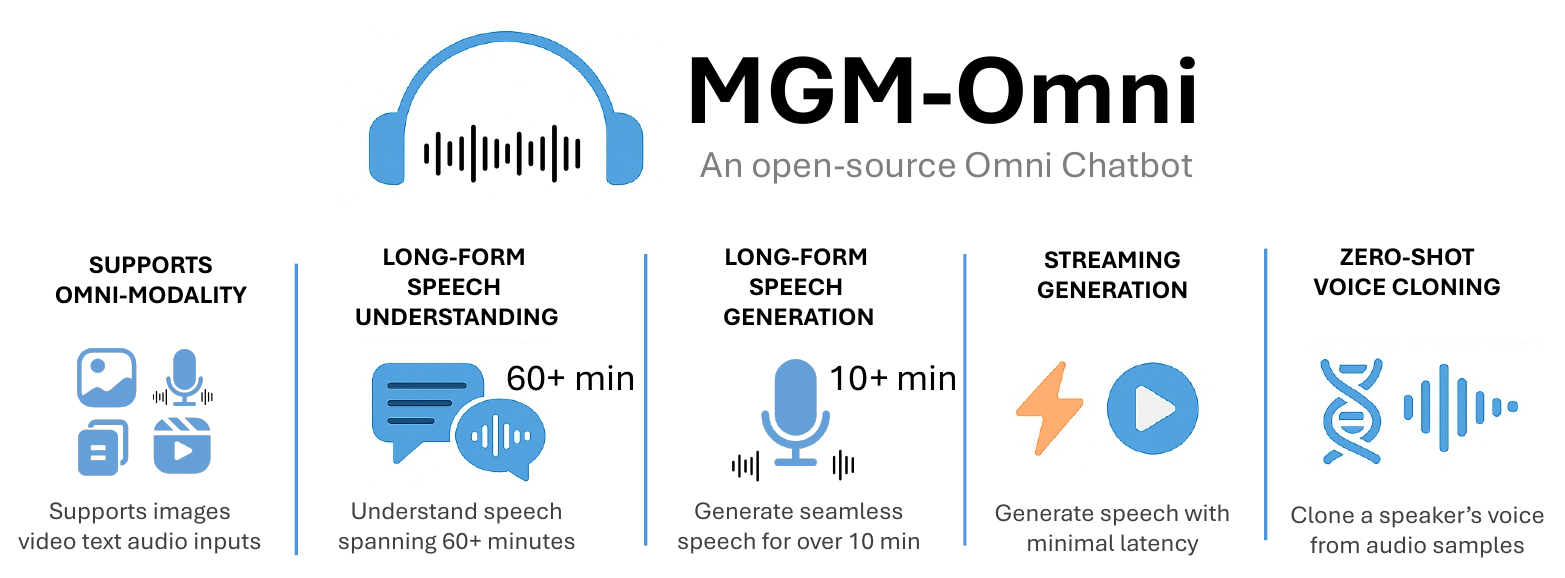}
\caption{MGM-Omni is an advanced Omni LLM for omnimodal understanding, long-form understanding, long-form speech generation and zero-shot voice clone. It can comprehend audio inputs exceeding 60 minutes and produce consistent, high-quality speech outputs longer than 10 minutes.}
\label{mgm-omni}
\end{center}
\end{figure}

\begin{abstract}
We present MGM-Omni, a unified Omni LLM for omni-modal understanding and expressive, long-horizon speech generation. Unlike cascaded pipelines that isolate speech synthesis, MGM-Omni adopts a "brain-mouth" design with a dual-track, token-based architecture that cleanly decouples multimodal reasoning from real-time speech generation. This design enables efficient cross-modal interaction and low-latency, streaming speech generation.
For understanding, a unified training strategy coupled with a dual audio encoder design enables long-form audio perception across diverse acoustic conditions.
For generation, a chunk-based parallel decoding scheme narrows the text speech token-rate gap, accelerating inference and supporting streaming zero-shot voice cloning with stable timbre over extended durations. Compared to concurrent work, MGM-Omni achieves these capabilities with markedly data-efficient training.
Extensive experiments demonstrate that MGM-Omni outperforms existing open source models in preserving timbre identity across extended sequences, producing natural and context-aware speech, and achieving superior long-form audio and omnimodal understanding.
MGM-Omni establishes an efficient, end-to-end paradigm for omnimodal understanding and controllable, personalised long-horizon speech generation.
\end{abstract}

\section{Introduction}
\label{intro}
\vspace{-2mm}

The evolution of large language models (LLMs) from purely text-based systems~\citep{chatgpt, llama} to multimodal frameworks has marked a significant paradigm shift in artificial intelligence. Vision language models (VLMs) such as LLaVA, GPT-4V, and Gemini~\citep{llava, gpt4, gemini} have demonstrated remarkable capabilities in understanding and processing visual information, effectively bridging the gap between vision and language. Audio serves as a bridge between humans and AI. However, integration of audio, particularly understanding and generating long-form and expressive audio, remains a significant challenge in multimodal systems. Most existing approaches are vision-centric, treating audio as a secondary input modality and relying on separate, cascaded text-to-speech (TTS) systems for generation~\citep{van2016wavenet, seedtts,cosyvoice2}. These methods exhibit critical shortcomings, including limited capability to process and understand extended audio sequences, high latency in audio synthesis, and degraded vocal timbre consistency over long durations.

\begin{wraptable}{r}{0.7\linewidth}
\vspace{-\baselineskip}
\centering
\resizebox{\linewidth}{!}{
\begin{tabular}{lcccccc}
\toprule
Model & VU & AU & LAU & SG & LSG & VC \\ \midrule
CosyVoice2~\citep{cosyvoice2}      &    &    &     & \checkmark &     & \checkmark \\
Higgs-Audio-v2~\citep{higgs-audio} &    &    &     & \checkmark & \checkmark & \checkmark \\
Qwen2.5-VL~\citep{qwen25vl}        & \checkmark &    &     &    &     &    \\
Qwen2.5-Omni~\citep{qwen25omni}    & \checkmark & \checkmark &     & \checkmark &     &    \\
Lyra~\citep{lyra}                  & \checkmark & \checkmark & \checkmark & \checkmark &     &    \\
\rowcolor{mygray} MGM-Omni                           & \checkmark & \checkmark & \checkmark & \checkmark & \checkmark & \checkmark \\ \bottomrule
\end{tabular}%
}
\caption{\small \textbf{Function comparison.} VU, AU, LAU, SG, LSG, and VC denote visual understanding, audio understanding, long audio understanding, speech generation, long speech generation, and zero-shot voice cloning.}
\end{wraptable}

The integration of audio in multimodal systems is hindered by the disparity between audio and text modalities. Audio token sequences are significantly more extensive and operate at a finer temporal resolution compared to their corresponding text token sequences~\citep{van2016wavenet, shen2018natural}. This disparity creates three challenges. First, existing systems lack robust long-form audio understanding, struggling to maintain contextual coherence and semantic accuracy across extended audio inputs. Second, in generation, a one-to-many alignment problem complicates mapping semantic words or units to long acoustic sequences, leading to misaligned prosody and unnatural pacing in long-form speech. Third, the autoregressive generation process is prone to error accumulation, where minor inaccuracies cascade, degrading timbre consistency and audio quality. Despite recent progress~\citep{stepaudio, qwen25omni, moss-ttsd, higgs-audio}, these systems do not address the intertwined issues of long-form audio understanding, alignment, and generation.

To address these limitations, we introduce MGM-Omni, an Omni LLM that unifies vision, language, and audio in an end-to-end framework for seamless, low-latency multimodal understanding and generation.
MGM-Omni adopts a dual-track architecture, separating multimodal reasoning (MLLM, the brain) from speech synthesis (SpeechLM, the mouth), enabling efficient cross-modal processing and real-time audio generation.
For audio understanding, we employ a dual-encoder design that fuses acoustic and semantic features, with unified training enabling unified inference across short and long audio.
For speech generation, we introduce Chunk-Based Parallel Decoding, which mitigates the token-rate gap between text and speech by segmenting text and predicting multiple speech tokens in parallel. This improves multimodal alignment, reduces long-sequence error accumulation and boosts inference speed by up to 3x.
Trained on approximately 400k hours of audio, MGM-Omni supports zero-shot voice cloning from any personalized reference voice.
Furthermore, we propose Long-TTS-Eval, a benchmark that systematically assesses long-form speech generation capability.
Consequently, MGM-Omni delivers zero-shot voice cloning and expressive, personalized long-horizon speech, maintaining timbre consistency and robust text-speech alignment across extended contexts. 
Our main contributions are threefold:
\begin{itemize}
\item We propose MGM-Omni, an Omni LLM featuring a novel dual-track design that unifies omni-modal understanding and expressive speech generation, moving beyond cascaded systems.
\item We introduce a Chunk-Based Parallel Decoding mechanism that mitigates the token-rate mismatch between text and speech, enabling efficient, high-fidelity, and context-aware long-form audio synthesis with customized voice.
\item Through extensive experiments, we demonstrate that MGM-Omni significantly outperforms existing methods in long audio understanding, and achieves leading performance in zero-shot voice cloning and natural, context-aware long-form speech generation.
\end{itemize}

\section{Related Work}
\label{related}

\paragraph{Multi-modal Large Language Models.}
The advent of large language models (LLMs)~\citep{chatgpt, llama} has revolutionized natural language processing, paving the way for multimodal extensions that integrate diverse data modalities such as text, image, video and audio~\citep{qwen25vl, qwen25omni, llamavid, mgm, llava, segzero}. Early multimodal models centered on vision-language alignment via contrastive learning. CLIP~\citep{clip} demonstrated the efficacy of zero-shot image classification through joint embedding spaces. Building on this foundation, vision language models (VLMs) like Flamingo, LLaVA and MiniGPT-4~\citep{flamingo,llava,minigpt4} adapted frozen visual encoders (e.g., CLIP-ViT) to instruction-tuned LLMs to enable general-purpose multimodal understanding. Subsequent works such as Mini-Gemini~\citep{mgm}, the LLaVA series~\citep{llava1.5,llava-onevision}, and the Qwen-VL series~\citep{qwen2vl,qwen25vl} further advance VLMs with high-resolution image comprehension, video understanding and visual grounding.
Despite this progress, most MLLMs remain vision-centric, with limited support for audio modalities. Recent efforts~\citep{vita1.5, lyra, qwen25omni} start to incorporate audio into MLLMs, but still struggle with understanding and generation of long-form audio, and cannot control the timbre of generated speech.
MGM-Omni address these limitations with a dual-track, token-based architecture that natively fuses language and audio, enabling omni-modal understanding and expressive, controllable long-form audio generation.

\paragraph{Speech Generation.}
In recent years, driven by the emergence of large language models (LLMs) and large-scale speech-text pre-training, zero-shot text-to-speech generation (TTS) has advanced markedly~\citep{seedtts,cosyvoice2,higgs-audio}.
CosyVoice2~\citep{cosyvoice2} builds a TTS system with chunk-aware flow matching and LLMs, enabling streaming multilingual speech synthesis with zero-shot voice cloning. Qwen2.5-Omni~\citep{qwen25omni} incorporates this design with a thinker-talker pipeline for end-to-end perception and generation across text, images, audio, and video. However, these systems still struggle with long-form speech generation. More recent efforts such as MOSS-TTSD~\citep{moss-ttsd} and Higgs-Audio-v2~\citep{higgs-audio} support expressive bilingual dialogue generation with personalized voice, yet challenges remain in maintaining timbre consistency over long sequences, ensuring real-time cross-modal fidelity, and achieving low latency.
MGM-Omni addresses this issue via a chunk-based parallel decoding approach, enabling expressive long-form speech generation with consistent timbre and low latency.

\section{MGM-Omni}
\label{method}

\begin{figure}[t]
\begin{center}
\includegraphics[width=1.0\textwidth]{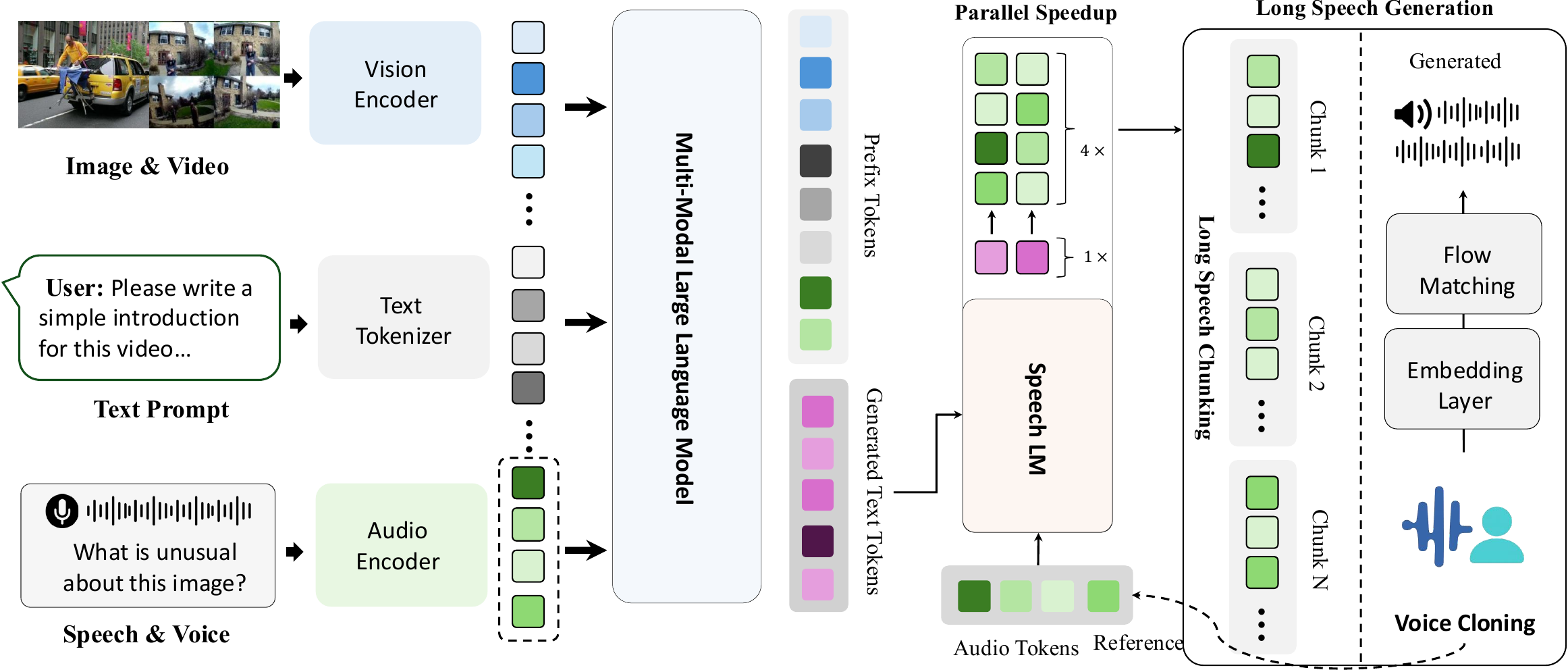}
\caption{\textbf{The overview of MGM-Omni.} MGM-Omni decouples omni-modal understanding and speech generation into an MLLM and a SpeechLM. The MLLM processes text, images, video, and audio to produce text, while the SpeechLM generates speech from the MLLM's output in real time.}
\label{framework}
\end{center}
\end{figure}

MGM-Omni is capable of processing text, images, video and speech, and can generate both textual and spoken outputs. To support high-quality, long-form speech synthesis without compromising the efficiency and effectiveness of omnimodal understanding and text generation, MGM-Omni decouples multimodal understanding and speech generation into two components: MLLM, serving as the "brain" for multimodal understanding and text generation, and SpeechLM, serving as the "mouth" for real-time speech generation.
For input in different modalities, we employ modality-specific encoders to extract features, which are subsequently passed to the MLLM. The MLLM generates text tokens and passes them to SpeechLM, which produces speech tokens in real-time via a Chunk-Based Parallel Decoding strategy. These speech tokens are further converted into Mel-spectrograms through a flow-matching model~\citep {flow-matching}, and the final audio is synthesized using a vocoder.
The overall framework is illustrated in Figure~\ref{framework}.

\subsection{Omni Understanding}

MGM-Omni is built upon Qwen2.5-VL~\citep{qwen25vl}, a state-of-the-art open-source Vision-Language Model (VLM) that supports image and video understanding with a native-resolution ViT~\citep{navit}. Based on Qwen2.5-VL, MGM-Omni attempts to extend towards Omni-Understanding, especially by incorporating audio understanding capabilities.

\paragraph{Dual Audio Encoder.}
MGM-Omni adopts a dual audio encoder design to capture both acoustic and semantic audio features.
The primary encoder, Qwen2-Audio~\citep{qwen2audio}, is an audio encoder continually trained on Whisper-large-v3~\citep{whisper} for enhanced general sound perception. 
To strengthen semantic understanding, especially for Chinese speech, we incorporate Belle-Whisper-large-v3~\citep{belle}, another Whisper-based encoder specialized in Chinese speech recognition.
This dual encoder setup yields two complementary representations: the main audio feature $X_{\text{main}}$ and the auxiliary audio feature $X_{\text{aux}}$.

\paragraph{Information Mining.}
To effectively integrate these complementary features, we design an audio information mining approach inspired by Mini-Gemini~\citep{mgm}.
Specifically, $X_{\text{main}}$ serves as the query $Q\in\mathbb{R}^{N\times C}$, while $X_{\text{aux}}$ provides the key-value pair: $K\in\mathbb{R}^{N\times C}$ and $V\in\mathbb{R}^{N\times C}$, allowing the model to retrieve semantically relevant cues from $X_{\text{aux}}$ under the guidance of $X_{\text{main}}$. Formally, information mining can be defined as:
\begin{equation}
    T_A= {\mathrm {MLP}}(Q + {\mathrm {Softmax}}(\phi(Q) \times \phi(K)^\top) \times \phi(V)),
\end{equation}
where $\phi$ denotes a projection layer and MLP represents a multi-layer perceptron. This approach enhances the audio representation by making it both acoustically and semantically aware, yielding enhanced audio tokens $T_A$ for subsequent LLM processing.

\paragraph{Training Strategy.}
Following Lyra~\citep{lyra}, we build a two-stage training pipeline to integrate audio understanding capabilities. In the first stage, we conduct audio-to-text pre-training to align the audio encoder to LLM. In the second stage, we perform unified omni-modal training. The first stage primarily uses audio transcription data, while the second stage comprises audio transcription, audio QA, audio-instruct VQA, and text instruction tuning data. This training strategy enables omni-cognition and robust cross-modal reasoning.

\paragraph{Omni Length Understanding.}
MGM-Omni aims to support both long and short sequence input. However, training with sequences of diverse lengths is inefficient: large batch sizes cause long sequence samples to run out of memory, while small sizes waste memory on short sequence samples. To address this issue, we propose a unified training pipeline. First, we group audio of similar lengths into the same batch. Second, we dynamically adjust the batch size, smaller for long-context inputs and larger for short-context inputs. This strategy significantly improves training efficiency.

\begin{figure}[t]
\begin{center}
\includegraphics[width=1.0\textwidth]{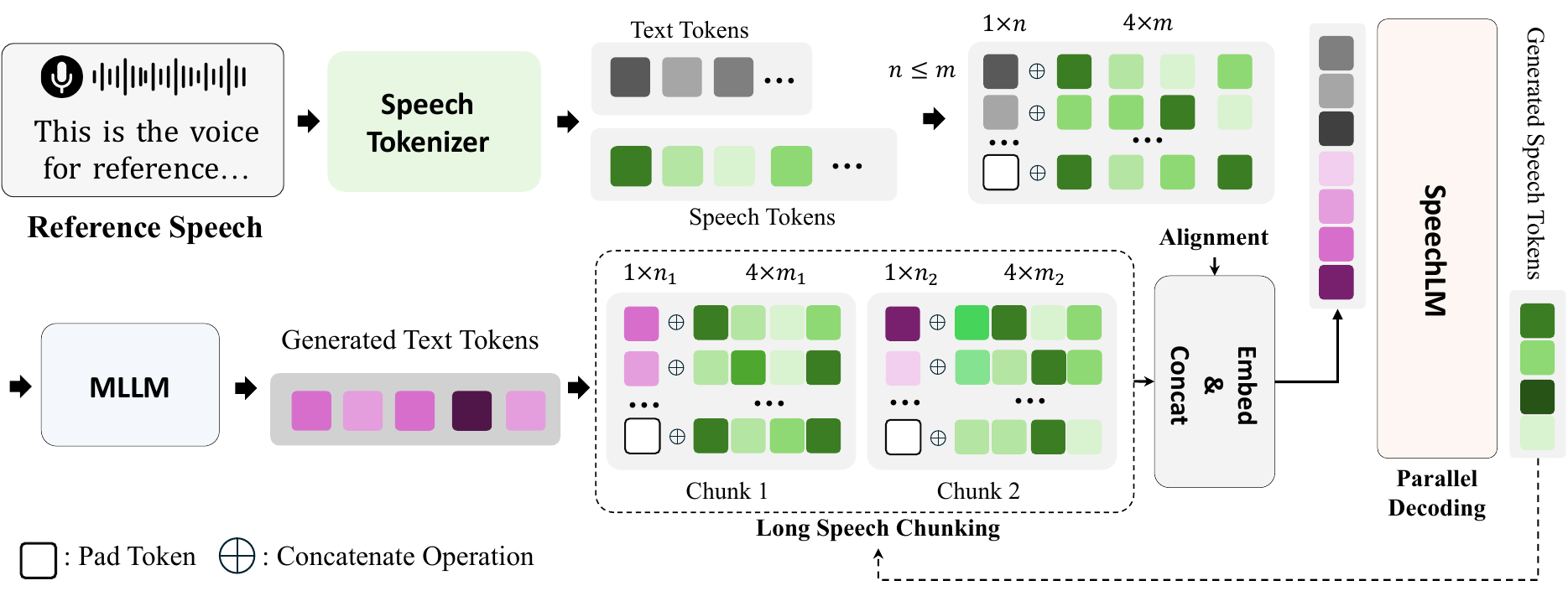}
\vspace{-2mm}
\caption{\textbf{The overview of SpeechLM in MGM-Omni.} Conditioned on MLLM-generated text and the reference audio clip, SpeechLM generate speech with Chunk-based Parallel Decoding.}
\label{speechlm}
\end{center}
\end{figure}

\subsection{Omni Generation}

MGM-Omni can generate both long-form text and speech. The textual output is autoregressively produced by the Omni-MLLM. The generated text, together with the personalize reference audio, is subsequently served as the conditioning for SpeechLM to synthesize speech via a Chunk-based Parallel Decoding method. The overall speech generation pipeline is depicted in Figure~\ref{speechlm}.

\paragraph{Speech Generation.} 
SpeechLM takes text tokens from Omni-MLLM as input and generates speech tokens in an autoregressive manner. It is initialized from the Qwen3~\citep{qwen3} language model, with an additional TTS-Adapter appended to its output. TTS-Adapter consists of six randomly initialized Qwen3 blocks, designed to transform text representations into speech representations.
The speech tokens produced by SpeechLM are then converted into Mel-spectrograms through a Flow-Matching model, and finally synthesized into audio via HiFi-GAN~\citep{hifigan} vocoder. We used the flow-matching model from CosyVoice2~\citep{cosyvoice2}, which supports chunk-aware streaming decoding.

\paragraph{Speech Tokenizer.}
We employ the CosyVoice2 finite scalar quantization~(FSQ) speech tokenizer to obtain discrete speech representations for speech generation. The tokenizer operates at a rate of 25 Hz, meaning that 25 tokens represent one second of audio. In comparison, humans typically express only two or three words per second. This discrepancy highlights that for a given utterance, the number of speech tokens is substantially larger than the number of text tokens. This leads to the following two issues:
\begin{itemize}
    \item As the length of the speech increases, the gap between text and speech tokens widens, weakening their correlation and degrading the quality of long-form generation.
    \item The much higher number of speech tokens compared to text tokens slows inference and harms streaming efficiency.  
\end{itemize}
To address these two challenges, we propose a Chunk-Based Parallel Decoding for efficient long-form speech generation.

\begin{figure}[t]
\begin{center}
\includegraphics[width=1.0\textwidth]{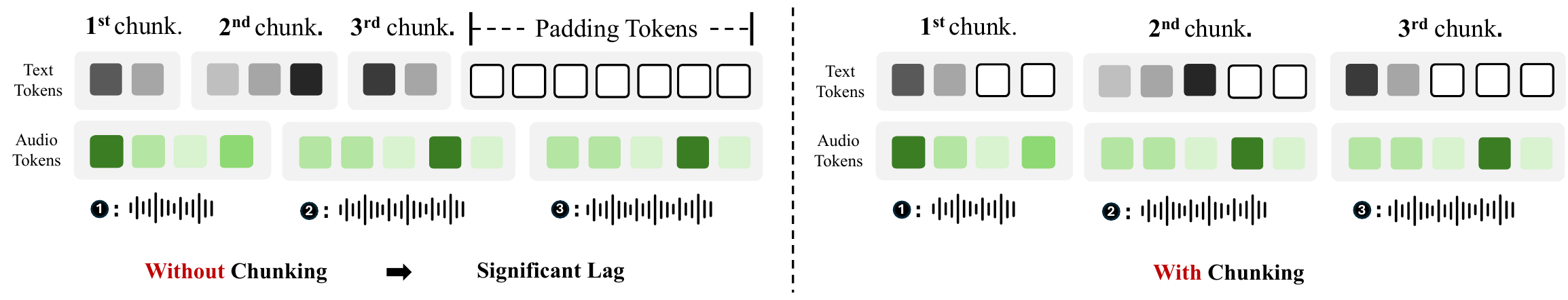}
\caption{\textbf{Decoding comparison.} Chunk-based decoding narrows the gap between text and corresponding speech, enabling long-form speech generation.}
\label{chunking}
\end{center}
\end{figure}

\paragraph{Chunk-based Decoding.}
To improve text-speech alignment in long-form speech generation, we introduce Chunk-based Decoding for speech token generation.
As shown in Figure~\ref{chunking}, the input text is divided into smaller chunks that are sequentially processed by SpeechLM, with each chunk producing a corresponding speech segment. 
During decoding, we adopt a token delay strategy: speech token generation within a chunk is initiated only after the first four text tokens, which are replaced by padding tokens in the speech sequence. This design ensures that every speech token is aligned with its corresponding text token while avoiding early mis-synchronization.
By reducing the alignment distance between modalities, Chunk-based Decoding enhances cross-modal correspondence and improves the robustness of long-form speech synthesis.
In contrast to naive segmentation methods, our approach preserves both the previously generated text and speech as context, thereby maintaining global fluency and coherence in the final output.
Notably, Chunk-based Decoding is highly compatible with our dual-track "brain-mouth" design, preserving omnimodal understanding and text generation speed while improving speech synthesis quality.

\paragraph{Parallel Decoding.}
To improve efficiency, we introduce a parallel decoding strategy for efficient speech token generation.
Specifically, we extend the vocabulary so that SpeechLM can decode both modalities in a single step. Let $V_{\text{text}}$ denote the text vocabulary, $V_{\text{speech}}$ denote the speech tokenizer vocabulary, and $k$ denote the parallel size. The extended vocabulary size is thus defined as:
\begin{equation}
    |V| = |V_{\text{text}}| + k |V_{\text{speech}}|.
\end{equation}
For speech tokenization, the input for each decoding step $t$ consists of one text token $x_t$ and $k$ speech tokens $\{s_t^1, s_t^2, \dots, s_t^k\}$. We use $f(\cdot)$ to denote the embedding function, and the hidden features $h_{t}^{\text{in}}$ for LLM input can be averaged as:
\begin{equation}
    h_{t}^{\text{in}} = \frac{1}{k+1} \left( f(x_t) + \sum_{i=1}^{k} f(s_t^i) \right).
\end{equation}
For speech detokenization, we employ a TTS-Adapter to project the LLM output hidden state $h_{t}^{\text{out}}$ into the speech representation space, after which the lm\_head predicts the next set of speech tokens:
\begin{equation}
    \{\hat{s}_{t+1}^1, \dots, \hat{s}_{t+1}^k\} = \text{lm\_head}\!\left(\text{TTS-Adapter}(h_{t}^{\text{out}})\right).
\end{equation}
While parallel decoding is commonly used with RVQ speech tokenizers~\citep{mini-omni, moss-ttsd}, it is rarely applied to FSQ speech tokenizers. We found that using parallel decoding with FSQ speech tokenizers not only maintains speech synthesis performance but also significantly improves efficiency. Additionally, it further shortens the distance between text and speech tokens, enhancing their correlation.

\subsection{Omni Voice}

MGM-Omni is capable of generating long-form speech in any personalized voice. To enable this capability, we carefully designed both the data pipeline and the training strategy.

\paragraph{Training Data.}
To enable zero-shot voice cloning, we collected a large-scale dataset, including around 300k hours of raw speech data and approximately 100k hours of TTS-synthesized speech in Chinese and English.
The raw speech portion of our corpus incorporates diverse open-source datasets, including Emilia Dataset~\citep{emilia}, Libri-heavy~\citep{libriheavy}, Common Voice~\citep{common-voice}, and Aishell series~\citep{aishell1, aishell2, aishell3}.
We constructed a dataset for TTS synthesis by sampling Chinese conversations from Belle-10M~\citep{belle} and English conversations from Lamini-Instruct~\citep{lamini-lm}. We uniformly sampled 900k Chinese and 700k English conversations based on length. As these datasets are somewhat outdated, we enhanced the text quality by regenerating all responses using Qwen2.5-72B~\citep{qwen2.5}. Subsequently, we synthesized audio from these refined conversations using megatts3~\citep{megatts3}. For each sample, we randomly select a reference voice from the provided set of pre-processed reference audio.

\paragraph{Pre-training.}
The SpeechLM consists of a pre-trained Qwen3~\citep{qwen3} LLM paired with a randomly initialized TTS-Adapter.
The model is trained to generate speech from given text and reference audio through a next speech token prediction objective.
The goal of the pre-training stage is to align the speech and text modalities. At this stage, the parameters of the pre-trained Qwen3 LLM remain frozen, while only the TTS-Adapter is updated. Both raw and synthesized speech data are leveraged in pre-training to ensure robustness across diverse speaker timbres.

\paragraph{Post-training.}
The post-training phase aims to enhance SpeechLM's capacity for fluent and accurate speech generation. During this phase, the parameters of both the LLM and the TTS-Adapter are jointly optimized with different learning rates. The TTS-Adapter is trained at a rate five times higher than that of the LLM. The training corpus is primarily composed of high-fidelity TTS-synthesised speech, supplemented with a smaller portion of raw speech data.

\section{Experiments}
\label{experiments}

\subsection{Main Properties}
\vspace{-2mm}

In this section, we present a comprehensive evaluation covering audio understanding, omni-modality understanding, and speech generation, to demonstrate the main properties of MGM-Omni, with particular emphasis on its capacity for long audio understanding and long audio generation and zero-shot voice cloning.

\begin{table}[t]
\centering
\resizebox{0.9\columnwidth}{!}{%
\renewcommand{\arraystretch}{1.0}%
\Large
\begin{tabular}{lccccc}
\toprule
\multicolumn{1}{l}{\multirow{2}{*}{Model}} & \multicolumn{2}{c}{LibriSpeech Test} & \multicolumn{2}{c}{CommonVoice} & AISHELL \\ \cmidrule{2-6} 
\multicolumn{1}{c}{} & clean WER ${\textcolor[rgb]{0.8,0.1,0.4}{\downarrow}}$ & other WER ${\textcolor[rgb]{0.8,0.1,0.4}{\downarrow}}$ & EN WER ${\textcolor[rgb]{0.8,0.1,0.4}{\downarrow}}$ & ZH CER ${\textcolor[rgb]{0.8,0.1,0.4}{\downarrow}}$ & CER ${\textcolor[rgb]{0.8,0.1,0.4}{\downarrow}}$ \\ \midrule
\multicolumn{6}{l}{\textbf{\textcolor{gray}{Audio LLMs}}} \\ \midrule
Whisper-large-v3~\citep{whisper} & 1.8 & 3.6 & 9.3 & 12.8            &  \\
Qwen2-Audio~\citep{qwen2audio}   & \textbf{1.3} & \textbf{3.4} & \textbf{8.6} & \textbf{5.2} &  \\ \midrule
\multicolumn{6}{l}{\textbf{\textcolor{gray}{Omni LLMs}}} \\ \midrule
Mini-Omni2~\citep{mini-omni2}     & 4.7 & 9.4 &  &  &  \\
Lyra~\citep{lyra}   & 2.0 & 4.0 &  &  &  \\
VITA-1.5~\citep{vita1.5}          & 3.4 & 7.5 &  &  & 2.2 \\
Ola~\citep{ola}     & 1.9 & 4.3 &  &  &   \\
Qwen2.5-Omni~\citep{qwen25omni}   & 1.6 & 3.5 & \textbf{7.6} & 5.2 &  \\ \midrule
\rowcolor{mygray}
MGM-Omni-7B        & 1.7 & 3.6 & 8.8 & 4.5 & 1.9 \\
\rowcolor{mygray}
MGM-Omni-32B       & \textbf{1.5} & \textbf{3.2} & 8.0 & \textbf{4.0} & \textbf{1.8} \\ \bottomrule
\end{tabular}%
}
\caption{\textbf{Omni-comparison on ASR benchmarks.} We use Common-Voice, LibriSpeech and AISHELL to evaluate the ASR capability on Chinese and English.}
\label{tab:speech_understanding}
\end{table}

\begin{table}[t]
\centering
\resizebox{0.9\columnwidth}{!}{%
{
\renewcommand{\arraystretch}{0.9}%
\tiny
\begin{tabular}{lccccc}
\toprule
{Model} & {Speech ${\textcolor[rgb]{0.1,0.7,0.4}{\uparrow}}$} & {Sound ${\textcolor[rgb]{0.1,0.7,0.4}{\uparrow}}$} & {Music ${\textcolor[rgb]{0.1,0.7,0.4}{\uparrow}}$} & {Mix ${\textcolor[rgb]{0.1,0.7,0.4}{\uparrow}}$} & {Average ${\textcolor[rgb]{0.1,0.7,0.4}{\uparrow}}$} \\
\midrule
\multicolumn{6}{l}{\textbf{\textcolor{gray}{Audio LLMs}}} \\
\midrule
SpeechGPT~\citep{speechgpt}    & 1.6 & 1.0 & 1.0 & 4.1 & 1.9 \\
SALMONN~\citep{salmonn}        & 6.2 & 6.3 & 6.0 & 6.1 & 6.1 \\
Qwen2-Audio~\citep{qwen2audio} & \textbf{7.2} & \textbf{7.0} & \textbf{6.8} & \textbf{6.8} & \textbf{6.9} \\
\midrule
\multicolumn{6}{l}{\textbf{\textcolor{gray}{Omni LLMs}}} \\
\midrule
LLaMA-Omni~\citep{llama-omni}    & 5.2 & 5.3 & 4.3 & 4.0 & 4.7 \\
Mini-Omni2~\citep{mini-omni}     & 3.6 & 3.5 & 2.6 & 3.1 & 3.2 \\
IXC2.5-OmniLive~\citep{internlm} & 1.6 & 1.8 & 1.7 & 1.6 & 1.7 \\
VITA-1.5~\citep{vita1.5}         & 4.8 & 5.5 & 4.9 & 2.9 & 4.5 \\
Qwen2.5-Omni~\citep{qwen25omni}  & 6.8 & 5.7 & 4.8 & 5.4 & 5.7 \\
Ola~\citep{ola}    & \textbf{7.3} & 6.4 & 5.9 & 6.0 & 6.4 \\ \midrule
\rowcolor{mygray}
MGM-Omni-7B     & \textbf{7.3} & \textbf{6.5} & \textbf{6.3} & 6.1 & \textbf{6.5} \\
\rowcolor{mygray}
MGM-Omni-32B    & 7.1 & \textbf{6.5} & 6.2 & \textbf{6.2}& \textbf{6.5}\\
\bottomrule
\end{tabular}
}
}
\caption{\textbf{Omni-comparison on Audio QA benchmarks.} We use AIR-Bench for audio QA evaluation. The scores are evaluated by gpt-4-0125-preview.}
\label{tab:audio_qa}
\end{table}

\subsubsection{Audio Understanding}
\vspace{-1mm}

\paragraph{Short Audio Understanding.}
We compare the audio understanding ability (audio $\rightarrow$ text) of MGM-Omni against leading Audio and Omni LLMs on two primary tasks: automatic speech recognition (ASR) and general audio QA.
First, we evaluate the ASR ability on LibriSpeech~\citep{libri-speech}, CommonVoice~\citep{common-voice} and AISHELL~\citep{aishell1}. As shown in Table~\ref{tab:speech_understanding}, MGM-Omni delivers competitive or superior performance for both English and Chinese ASR. In particular, with dual audio encoder, MGM-Omni achieves 4.0 CER on CommonVoice (ZH) and 1.8 CER on AISHELL, surpassing leading audio and Omni LLMs.
For general audio understanding, we evaluate audio QA on AIR-Bench~\citep{airbench}, a comprehensive benchmark covering speech, sound, and music inputs.
As summarized in Table~\ref{tab:audio_qa}, MGM-Omni outperforms all open source Omni LLMs, including Qwen2.5-Omni~\citep{qwen25omni}.

\begin{figure}[h]
\begin{center}
\includegraphics[width=1.0\textwidth]{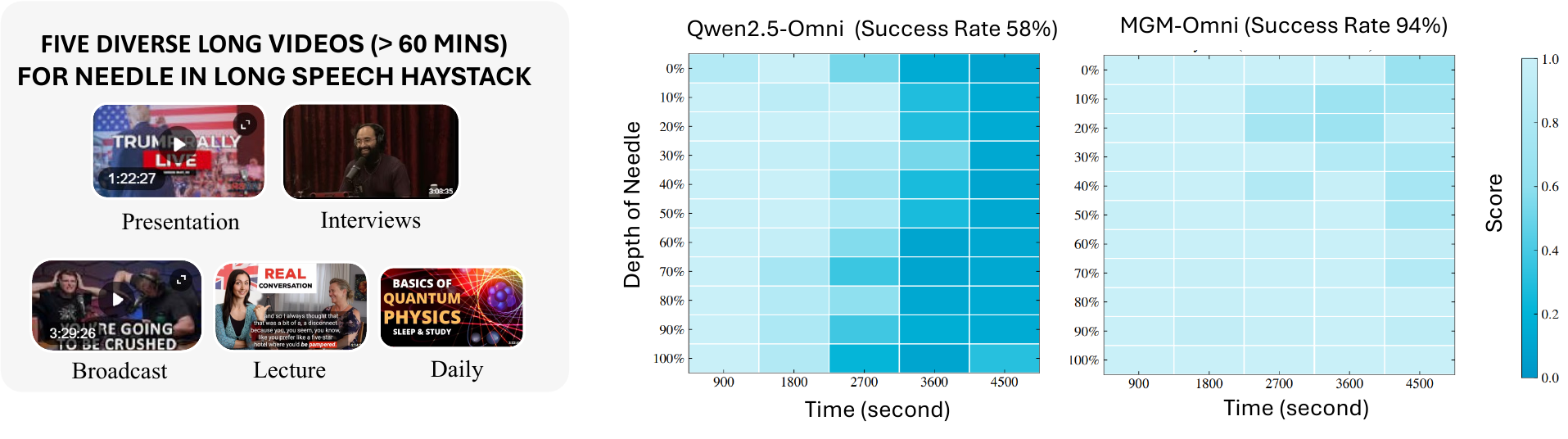}
\caption{\textbf{Omni-comparison for Long-form Audio.} We adopt a needle-in-the-haystack evaluation and report the average success rate across five materials.}
\label{long-audio}
\end{center}
\end{figure}

\vspace{-2mm}
\paragraph{Long Audio Understanding.}
Unlike many open-source Audio and Omni LLMs, MGM-Omni is capable of processing audio inputs exceeding one hour in length. To evaluate its ability on long-form audio understanding, we conducted a needle-in-the-haystack test. As illustrated in Figure~\ref{long-audio}, MGM-Omni successfully handles audio inputs of up to 4,500 seconds, significantly outperforming Qwen2.5-Omni~\citep{qwen25omni}. The success rate is averaged over five diverse long-form audio. Moreover, we provide quantitative comparison in Figure~\ref{fig:long-asr} in the appendix.

\begin{table}[t]
\centering
\resizebox{\columnwidth}{!}{%
\begin{tabular}{lcccc}
\toprule
{Model} & TextVQA-Speech ${\textcolor[rgb]{0.1,0.7,0.4}{\uparrow}}$ & DocVQA-Speech ${\textcolor[rgb]{0.1,0.7,0.4}{\uparrow}}$ & ChartVQA-Speech ${\textcolor[rgb]{0.1,0.7,0.4}{\uparrow}}$ & AI2D-Speech ${\textcolor[rgb]{0.1,0.7,0.4}{\uparrow}}$ \\ \midrule
Intern-Omni-9B~\citep{internomni}  & 69.1 & 80.0 & 56.1 & 54.0 \\
Lyra-9B~\citep{lyra}               & 80.0 & 85.5 & 61.0 & 63.1 \\ \midrule
\rowcolor{mygray}
MGM-Omni-7B    & \textbf{81.7}  & 87.4 & 69.3 & 70.4 \\
\rowcolor{mygray}
MGM-Omni-32B   & 78.2 & \textbf{88.4} & \textbf{72.1} & \textbf{71.3} \\ \bottomrule
\end{tabular}%
}
\caption{\textbf{Omni-comparison on vision-speech benchmarks.} We convert the textual questions in multiple VQA benchmarks into synthesized speech to evaluate the multimodal understanding ability.}
\label{tab:vision-speech}
\end{table}

\subsubsection{Omni-Modality Understanding}
\vspace{-2mm}

MGM-Omni processes text, image, video, and audio inputs. Following Lyra~\citep{lyra}, we further evaluate its omni-modal understanding (multimodality $\rightarrow$ text) by comparing MGM-Omni against other omni-modal LLMs on several speech-instructed VQA benchmarks.
As shown in Table~\ref {tab:vision-speech}, MGM-Omni shows a strong ability to follow speech instructions.

\begin{table}[t]
  \centering

  \begin{subtable}[t]{\linewidth}
    \centering
    \resizebox{\linewidth}{!}{%
    \renewcommand{\arraystretch}{0.95}
    \tiny
    \begin{tabular}{lcccccc}
        \toprule
        \multicolumn{1}{l}{Model} & Size & EN WER ${\textcolor[rgb]{0.8,0.1,0.4}{\downarrow}}$ & EN SIM ${\textcolor[rgb]{0.1,0.7,0.4}{\uparrow}}$ & ZH CER ${\textcolor[rgb]{0.8,0.1,0.4}{\downarrow}}$ & ZH SIM ${\textcolor[rgb]{0.1,0.7,0.4}{\uparrow}}$ \\ \midrule
        CosyVoice2~\citep{cosyvoice2}      & 0.5B & 2.57 & 0.652 & 1.45 & 0.748 \\
        Qwen2.5-Omni-3B~\citep{qwen25omni} & 0.5B & 2.51 & 0.635 & 1.58 & 0.744 \\
        Qwen2.5-Omni-7B~\citep{qwen25omni} & 2B   & 2.33 & 0.641 & 1.42 & 0.754 \\
        Higgs-Audio-v2~\citep{higgs-audio} & 6B   & 2.44 & 0.677 & 1.66 & 0.743 \\ \midrule
        \rowcolor{mygray}
        MGM-Omni-TTS-0.6B & 0.6B & 2.48 & 0.670 & 1.42 & 0.750 \\
        \rowcolor{mygray}
        MGM-Omni-TTS-2B   & 2B   & 2.28 & 0.684 & 1.28 & 0.755 \\
        \rowcolor{mygray}
        MGM-Omni-TTS-4B   & 4B   & \textbf{2.22} & \textbf{0.686} & \textbf{1.18} & \textbf{0.758} \\ 
        \bottomrule
    \end{tabular}%
    }
    \caption{Zero-shot short TTS comparison of error rate and speaker similarity in Seed-TTS-Eval. For Qwen2.5-Omni, size indicates the talker module size.}
    \label{tab:tts}
  \end{subtable}

  \vspace{2mm}

  \begin{subtable}[t]{\linewidth}
    \centering
    \resizebox{\linewidth}{!}{%
    \renewcommand{\arraystretch}{1.1}
    \begin{tabular}{lcccccc}
        \toprule
        \multicolumn{1}{c}{Model} & Size & RTF ${\textcolor[rgb]{0.8,0.1,0.4}{\downarrow}}$ & EN WER ${\textcolor[rgb]{0.8,0.1,0.4}{\downarrow}}$ & ZH CER ${\textcolor[rgb]{0.8,0.1,0.4}{\downarrow}}$ & EN-hard WER ${\textcolor[rgb]{0.8,0.1,0.4}{\downarrow}}$ & ZH-hard CER ${\textcolor[rgb]{0.8,0.1,0.4}{\downarrow}}$ \\ \midrule
        CosyVoice2 (chunk)~\citep{cosyvoice2} & 0.5B & 0.34 & 14.80 & \textbf{5.27}  & 42.48 & 32.76 \\
        MOSS-TTSD-v0.5~\citep{moss-ttsd}     & 2B   & 0.23 & 8.69  & 6.82  & 62.61 & 62.97 \\
        Higgs-Audio-v2~\citep{higgs-audio}   & 6B   & 0.33 & 27.09 & 31.39 & 98.61 & 98.85 \\ \midrule
        \rowcolor{mygray}
        MGM-Omni-TTS-2B & 2B  & \textbf{0.19} & \textbf{4.98} & 5.58 & \textbf{26.26} & \textbf{23.58} \\
        \bottomrule
    \end{tabular}%
    }
    \caption{Long-form TTS comparison of error rate and inference speed in our Long-TTS-Eval.}
    \label{tab:long-tts}
  \end{subtable}

  \caption{\textbf{Omni-comparison on TTS benchmarks.} We evaluate short-form and long-form TTS using Seed-TTS-Eval (top) and Long-TTS-Eval (bottom).}
  \label{tab:combined-tts}
\end{table}

\subsubsection{Speech Generation}
\vspace{-2mm}

MGM-Omni supports long-form synthesis (exceeding 10 minutes) with customizable voices. Here, we assess the speech generation capabilities (text $\rightarrow$ speech) in both short- and long-form setting.

\vspace{-2mm}
\paragraph{Short Speech Generation.}
We evaluated MGM-Omni against state-of-the-art zero-shot TTS systems and Omni LLMs to assess the speech generation capabilities. As shown in Table~\ref{tab:tts}, MGM achieves lower error rates and higher speaker similarity than open-source TTS models and Omni LLMs on seed-tts-eval~\citep{seedtts}, demonstrating strong text-to-speech performance and robust zero-shot voice cloning.

\vspace{-2mm}
\paragraph{Long Speech Generation.}
Unlike many open-source Omni LLMs and TTS systems, MGM-Omni can generate over 10 minutes of speech in any personalized voice. Quantitative examples are shown in Figure~\ref{fig:long-tts} in the appendix.
For benchmark evaluation, most existing benchmarks only evaluate short clips, typically ranging from a few seconds to a few dozen seconds, leaving a gap in assessing long-form performance. Moreover, existing TTS benchmarks focus on normal text generation and do not cover more complex text, such as formulas, URLs, or classical Chinese poetry.
To address this, we introduce \textbf{Long-TTS-Eval}, a benchmark specifically designed to evaluate long-form text-to-speech generation systematically. We leave more detailed information about the benchmark to Section~\ref{sec:long-tts-eval} in the appendix. 

We compare MGM-Omni against two categories of open-source TTS systems: (1) Native long TTS models, represented by MOSS-TTSD-v0.5~\citep{moss-ttsd} and Higgs-Audio-v2~\citep{higgs-audio}. (2) Non-native models that extend via chunking, represented by CosyVoice2~\citep{cosyvoice2}. We report WER for English TTS, CER for Chinese TTS, and RTF for inference efficiency.
As shown in Table~\ref{tab:long-tts}, MGM-Omni achieves lower error rates across most speech generation scenarios, along with the lowest RTF.
It is worth noting that, MGM-Omni's two-stage training relies on less than 400k hours of audio, substantially fewer than the 1M or even 10M hours used in concurrent works.
This result demonstrates the efficiency, effectiveness, robustness and data efficiency of our model.

\begin{table}[t]
\centering
\begin{subtable}{0.32\linewidth}
\centering
\resizebox{\linewidth}{!}{%
\renewcommand{\arraystretch}{1.21}
\begin{tabular}{lcc}
\toprule
Audio Encoder  & EN WER ${\textcolor[rgb]{0.8,0.1,0.4}{\downarrow}}$ & ZH CER ${\textcolor[rgb]{0.8,0.1,0.4}{\downarrow}}$ \\
\midrule
 Qwen2-Audio  & 13.0 & 3.9 \\
 Belle-Whisper  & 21.7 & 5.0 \\
\rowcolor{mygray} Info Mining  & \textbf{9.1} & \textbf{3.5}\\
\bottomrule
\end{tabular}
}
\caption{Audio Encoder}
\label{tab:ablation_encoder}
\end{subtable}\hfill
\begin{subtable}{0.32\linewidth}
\centering
\resizebox{\linewidth}{!}{%
\renewcommand{\arraystretch}{1.21}
\begin{tabular}{cccc}
\toprule
Parallel & RTF & EN WER ${\textcolor[rgb]{0.8,0.1,0.4}{\downarrow}}$ & ZH CER ${\textcolor[rgb]{0.8,0.1,0.4}{\downarrow}}$ \\
\midrule
1 & 0.57 & \textbf{1.86} & \textbf{1.15} \\
2 & 0.32 & 2.02 & 1.23 \\
\rowcolor{mygray}
4 & \textbf{0.19} & 2.28 & 1.28 \\
\bottomrule
\end{tabular}
}
\caption{Parallel Decoding}
\label{tab:ablation_parallel}
\end{subtable}\hfill
\begin{subtable}{0.32\linewidth}
\centering
\resizebox{\linewidth}{!}{%
\renewcommand{\arraystretch}{1.38}
\Large
\begin{tabular}{cccc}
\toprule
Chunking & EN WER ${\textcolor[rgb]{0.8,0.1,0.4}{\downarrow}}$ & ZH CER ${\textcolor[rgb]{0.8,0.1,0.4}{\downarrow}}$ \\
\midrule
  & 31.84 & 8.97 \\
\rowcolor{mygray}
\checkmark & \textbf{4.98} & \textbf{5.64} \\
\bottomrule
\end{tabular}
}
\caption{Chunk-Based Decoding}
\label{tab:ablation_chunk}
\end{subtable}
\caption{\textbf{Ablation study.} We conduct ablation studies on the audio encoder, parallel decoding, and chunk-based decoding.}
\label{tab:ablation_study}
\end{table}

\subsection{Ablation Study}
\vspace{-2mm}

\paragraph{Audio Encoder.}
We ablate different audio encoder designs and evaluate on CommonVoice ASR~\citep{common-voice}. As shown in Table~\ref{tab:ablation_encoder}, incorporating both the Qwen2-Audio encoder~\citep{qwen2audio} and the Belle-Whisper-large-v3 encoder~\citep{belle} with information mining yields the best performance in audio understanding. Note that, compared with the final model, we do not use the long audio QA data here.

\vspace{-2mm}
\paragraph{Chunk-based Decoding.}
We evaluate long-form speech generation on our Long-TTS-Eval to assess the impact of chunk-based decoding. As shown in Table~\ref{tab:ablation_chunk}, removing chunk-based decoding leads to a substantially higher error rate, exceeding that of concurrent works. Given that concurrent methods typically use millions to tens of millions of hours of audio, we attribute MGM-Omni's data efficiency primarily to its use of chunk-based decoding.

\vspace{-2mm}
\paragraph{Parallel Decoding.}
We ablate the impact of parallel decoding by comparing both TTS performance and inference speed. TTS performance is measured on Seed-TTS-Eval~\citep{seedtts}, while inference speed is assessed using 16 Chinese and 16 English samples drawn from Long-TTS-Eval. We report the real-time factor (RTF) on a single H800 GPU to compare the inference speed.
As shown in Table~\ref{tab:ablation_parallel}, increasing the parallel size slightly raises the audio error rate but substantially accelerates inference by 3x. To balance quality and speed, we set the parallel size to 4. We anticipate that incorporating more advanced Multi-Token Prediction (MTP) techniques~\citep{deepseekv3} will further improve audio quality at larger parallel sizes.

\section{Conclusion}
\label{conclusion}
\vspace{-2mm}

We present MGM-Omni, a unified Omni LLM that supports long-form omnimodal understanding and robust long-duration speech generation with personalized voices. Its dual-track architecture separates multimodal reasoning (MLLM) from real-time speech synthesis (SpeechLM), enabling efficient cross-modal interaction within an end-to-end framework. For understanding, it employs a dual audio encoder that fuses acoustic and semantic cues, yielding robust long-form audio perception. For generation, we introduce Chunk-Based Parallel Decoding to bridge the token-rate gap between text and speech, enabling efficient, low-latency synthesis, while conditioning SpeechLM on reference audio to support zero-shot voice cloning with consistent timbre. Experiments show that MGM-Omni surpasses leading open source Omni LLMs in timbre consistency, context-aware speech, long audio comprehension, and omni-modal reasoning.

\bibliography{iclr2026_conference}

\begin{thebibliography}{56}
\providecommand{\natexlab}[1]{#1}
\providecommand{\url}[1]{\texttt{#1}}
\expandafter\ifx\csname urlstyle\endcsname\relax
  \providecommand{\doi}[1]{doi: #1}\else
  \providecommand{\doi}{doi: \begingroup \urlstyle{rm}\Url}\fi

\bibitem[Alayrac et~al.(2022)Alayrac, Donahue, Luc, Miech, Barr, Hasson, Lenc,
  Mensch, Millican, Reynolds, et~al.]{flamingo}
Jean-Baptiste Alayrac, Jeff Donahue, Pauline Luc, Antoine Miech, Iain Barr,
  Yana Hasson, Karel Lenc, Arthur Mensch, Katherine Millican, Malcolm Reynolds,
  et~al.
\newblock Flamingo: a visual language model for few-shot learning.
\newblock \emph{Advances in neural information processing systems},
  35:\penalty0 23716--23736, 2022.

\bibitem[Anastassiou et~al.(2024)Anastassiou, Chen, Chen, Chen, Chen, Chen,
  Cong, Deng, Ding, Gao, et~al.]{seedtts}
Philip Anastassiou, Jiawei Chen, Jitong Chen, Yuanzhe Chen, Zhuo Chen, Ziyi
  Chen, Jian Cong, Lelai Deng, Chuang Ding, Lu~Gao, et~al.
\newblock Seed-tts: A family of high-quality versatile speech generation
  models.
\newblock \emph{arXiv preprint arXiv:2406.02430}, 2024.

\bibitem[Ardila et~al.(2019)Ardila, Branson, Davis, Henretty, Kohler, Meyer,
  Morais, Saunders, Tyers, and Weber]{common-voice}
Rosana Ardila, Megan Branson, Kelly Davis, Michael Henretty, Michael Kohler,
  Josh Meyer, Reuben Morais, Lindsay Saunders, Francis~M Tyers, and Gregor
  Weber.
\newblock Common voice: A massively-multilingual speech corpus.
\newblock \emph{arXiv preprint arXiv:1912.06670}, 2019.

\bibitem[Bai et~al.(2025)Bai, Chen, Liu, Wang, Ge, Song, Dang, Wang, Wang,
  Tang, et~al.]{qwen25vl}
Shuai Bai, Keqin Chen, Xuejing Liu, Jialin Wang, Wenbin Ge, Sibo Song, Kai
  Dang, Peng Wang, Shijie Wang, Jun Tang, et~al.
\newblock Qwen2.5-vl technical report.
\newblock \emph{arXiv preprint arXiv:2502.13923}, 2025.

\bibitem[BELLEGroup(2023)]{belle}
BELLEGroup.
\newblock Belle: Be everyone's large language model engine.
\newblock \url{https://github.com/LianjiaTech/BELLE}, 2023.

\bibitem[{Boson AI}(2025)]{higgs-audio}
{Boson AI}.
\newblock {Higgs Audio V2: Redefining Expressiveness in Audio Generation}.
\newblock \url{https://github.com/boson-ai/higgs-audio}, 2025.
\newblock GitHub repository. Release blog available at
  \url{https://www.boson.ai/blog/higgs-audio-v2}.

\bibitem[Bu et~al.(2017)Bu, Du, Na, Wu, and Zheng]{aishell1}
Hui Bu, Jiayu Du, Xingyu Na, Bengu Wu, and Hao Zheng.
\newblock Aishell-1: An open-source mandarin speech corpus and a speech
  recognition baseline.
\newblock In \emph{2017 20th conference of the oriental chapter of the
  international coordinating committee on speech databases and speech I/O
  systems and assessment (O-COCOSDA)}, pp.\  1--5. IEEE, 2017.

\bibitem[Chu et~al.(2024)Chu, Xu, Yang, Wei, Wei, Guo, Leng, Lv, He, Lin, Zhou,
  and Zhou]{qwen2audio}
Yunfei Chu, Jin Xu, Qian Yang, Haojie Wei, Xipin Wei, Zhifang Guo, Yichong
  Leng, Yuanjun Lv, Jinzheng He, Junyang Lin, Chang Zhou, and Jingren Zhou.
\newblock Qwen2-audio technical report.
\newblock \emph{arXiv preprint arXiv:2407.10759}, 2024.

\bibitem[Comanici et~al.(2025)Comanici, Bieber, Schaekermann, Pasupat,
  Sachdeva, Dhillon, Blistein, Ram, Zhang, Rosen, et~al.]{gemini2.5}
Gheorghe Comanici, Eric Bieber, Mike Schaekermann, Ice Pasupat, Noveen
  Sachdeva, Inderjit Dhillon, Marcel Blistein, Ori Ram, Dan Zhang, Evan Rosen,
  et~al.
\newblock Gemini 2.5: Pushing the frontier with advanced reasoning,
  multimodality, long context, and next generation agentic capabilities.
\newblock \emph{arXiv preprint arXiv:2507.06261}, 2025.

\bibitem[Dehghani et~al.(2023)Dehghani, Mustafa, Djolonga, Heek, Minderer,
  Caron, Steiner, Puigcerver, Geirhos, Alabdulmohsin, et~al.]{navit}
Mostafa Dehghani, Basil Mustafa, Josip Djolonga, Jonathan Heek, Matthias
  Minderer, Mathilde Caron, Andreas Steiner, Joan Puigcerver, Robert Geirhos,
  Ibrahim~M Alabdulmohsin, et~al.
\newblock Patch n’pack: Navit, a vision transformer for any aspect ratio and
  resolution.
\newblock \emph{Advances in Neural Information Processing Systems},
  36:\penalty0 2252--2274, 2023.

\bibitem[Du et~al.(2018)Du, Na, Liu, and Bu]{aishell2}
Jiayu Du, Xingyu Na, Xuechen Liu, and Hui Bu.
\newblock Aishell-2: Transforming mandarin asr research into industrial scale.
\newblock \emph{arXiv preprint arXiv:1808.10583}, 2018.

\bibitem[Du et~al.(2024)Du, Wang, Chen, Shi, Lv, Zhao, Gao, Yang, Gao, Wang,
  et~al.]{cosyvoice2}
Zhihao Du, Yuxuan Wang, Qian Chen, Xian Shi, Xiang Lv, Tianyu Zhao, Zhifu Gao,
  Yexin Yang, Changfeng Gao, Hui Wang, et~al.
\newblock Cosyvoice 2: Scalable streaming speech synthesis with large language
  models.
\newblock \emph{arXiv preprint arXiv:2412.10117}, 2024.

\bibitem[Fang et~al.(2024)Fang, Guo, Zhou, Ma, Zhang, and Feng]{llama-omni}
Qingkai Fang, Shoutao Guo, Yan Zhou, Zhengrui Ma, Shaolei Zhang, and Yang Feng.
\newblock Llama-omni: Seamless speech interaction with large language models.
\newblock \emph{arXiv preprint arXiv:2409.06666}, 2024.

\bibitem[Fu et~al.(2024)Fu, Lin, Long, Shen, Dai, Zhao, Zhang, Dong, Li, Wang,
  et~al.]{vita1.5}
Chaoyou Fu, Haojia Lin, Zuwei Long, Yunhang Shen, Yuhang Dai, Meng Zhao, Yi-Fan
  Zhang, Shaoqi Dong, Yangze Li, Xiong Wang, et~al.
\newblock Vita: Towards open-source interactive omni multimodal llm.
\newblock \emph{arXiv preprint arXiv:2408.05211}, 2024.

\bibitem[Gao et~al.(2023)Gao, Li, Wang, Luo, Shi, Chen, Li, Zuo, Du, Xiao,
  et~al.]{funasr}
Zhifu Gao, Zerui Li, Jiaming Wang, Haoneng Luo, Xian Shi, Mengzhe Chen, Yabin
  Li, Lingyun Zuo, Zhihao Du, Zhangyu Xiao, et~al.
\newblock Funasr: A fundamental end-to-end speech recognition toolkit.
\newblock \emph{arXiv preprint arXiv:2305.11013}, 2023.

\bibitem[He et~al.(2024)He, Shang, Wang, Li, Gu, Hua, Liu, Yang, Li, Shi,
  et~al.]{emilia}
Haorui He, Zengqiang Shang, Chaoren Wang, Xuyuan Li, Yicheng Gu, Hua Hua, Liwei
  Liu, Chen Yang, Jiaqi Li, Peiyang Shi, et~al.
\newblock Emilia: An extensive, multilingual, and diverse speech dataset for
  large-scale speech generation.
\newblock In \emph{2024 IEEE Spoken Language Technology Workshop (SLT)}, pp.\
  885--890. IEEE, 2024.

\bibitem[Huang et~al.(2025)Huang, Wu, Wang, Yan, Hu, Feng, Tian, Shen, Li,
  Chen, et~al.]{stepaudio}
Ailin Huang, Boyong Wu, Bruce Wang, Chao Yan, Chen Hu, Chengli Feng, Fei Tian,
  Feiyu Shen, Jingbei Li, Mingrui Chen, et~al.
\newblock Step-audio: Unified understanding and generation in intelligent
  speech interaction.
\newblock \emph{arXiv preprint arXiv:2502.11946}, 2025.

\bibitem[Jiang et~al.(2025)Jiang, Ren, Li, Ji, Ye, Zhang, Jionghao, Yang, Zuo,
  Zhang, et~al.]{megatts3}
Ziyue Jiang, Yi~Ren, Ruiqi Li, Shengpeng Ji, Zhenhui Ye, Chen Zhang, Bai
  Jionghao, Xiaoda Yang, Jialong Zuo, Yu~Zhang, et~al.
\newblock Sparse alignment enhanced latent diffusion transformer for zero-shot
  speech synthesis.
\newblock \emph{arXiv preprint arXiv:2502.18924}, 2025.

\bibitem[Kang et~al.(2024)Kang, Yang, Yao, Kuang, Yang, Guo, Lin, and
  Povey]{libriheavy}
Wei Kang, Xiaoyu Yang, Zengwei Yao, Fangjun Kuang, Yifan Yang, Liyong Guo, Long
  Lin, and Daniel Povey.
\newblock Libriheavy: A 50,000 hours asr corpus with punctuation casing and
  context.
\newblock In \emph{ICASSP 2024-2024 IEEE International Conference on Acoustics,
  Speech and Signal Processing (ICASSP)}, pp.\  10991--10995. IEEE, 2024.

\bibitem[Kong et~al.(2020)Kong, Kim, and Bae]{hifigan}
Jungil Kong, Jaehyeon Kim, and Jaekyoung Bae.
\newblock Hifi-gan: Generative adversarial networks for efficient and high
  fidelity speech synthesis.
\newblock \emph{Advances in neural information processing systems},
  33:\penalty0 17022--17033, 2020.

\bibitem[Li et~al.(2024{\natexlab{a}})Li, Zhang, Guo, Zhang, Li, Zhang, Zhang,
  Zhang, Li, Liu, et~al.]{llava-onevision}
Bo~Li, Yuanhan Zhang, Dong Guo, Renrui Zhang, Feng Li, Hao Zhang, Kaichen
  Zhang, Peiyuan Zhang, Yanwei Li, Ziwei Liu, et~al.
\newblock Llava-onevision: Easy visual task transfer.
\newblock \emph{arXiv preprint arXiv:2408.03326}, 2024{\natexlab{a}}.

\bibitem[Li et~al.(2024{\natexlab{b}})Li, Wang, and Jia]{llamavid}
Yanwei Li, Chengyao Wang, and Jiaya Jia.
\newblock Llama-vid: An image is worth 2 tokens in large language models.
\newblock In \emph{European Conference on Computer Vision}, pp.\  323--340.
  Springer, 2024{\natexlab{b}}.

\bibitem[Li et~al.(2024{\natexlab{c}})Li, Zhang, Wang, Zhong, Chen, Chu, Liu,
  and Jia]{mgm}
Yanwei Li, Yuechen Zhang, Chengyao Wang, Zhisheng Zhong, Yixin Chen, Ruihang
  Chu, Shaoteng Liu, and Jiaya Jia.
\newblock Mini-gemini: Mining the potential of multi-modality vision language
  models.
\newblock \emph{arXiv preprint arXiv:2403.18814}, 2024{\natexlab{c}}.

\bibitem[Lipman et~al.(2022)Lipman, Chen, Ben-Hamu, Nickel, and
  Le]{flow-matching}
Yaron Lipman, Ricky~TQ Chen, Heli Ben-Hamu, Maximilian Nickel, and Matt Le.
\newblock Flow matching for generative modeling.
\newblock \emph{arXiv preprint arXiv:2210.02747}, 2022.

\bibitem[Liu et~al.(2024)Liu, Feng, Xue, Wang, Wu, Lu, Zhao, Deng, Zhang, Ruan,
  et~al.]{deepseekv3}
Aixin Liu, Bei Feng, Bing Xue, Bingxuan Wang, Bochao Wu, Chengda Lu, Chenggang
  Zhao, Chengqi Deng, Chenyu Zhang, Chong Ruan, et~al.
\newblock Deepseek-v3 technical report.
\newblock \emph{arXiv preprint arXiv:2412.19437}, 2024.

\bibitem[Liu et~al.(2023{\natexlab{a}})Liu, Li, Li, and Lee]{llava1.5}
Haotian Liu, Chunyuan Li, Yuheng Li, and Yong~Jae Lee.
\newblock Improved baselines with visual instruction tuning.
\newblock \emph{arXiv:2310.03744}, 2023{\natexlab{a}}.

\bibitem[Liu et~al.(2023{\natexlab{b}})Liu, Li, Wu, and Lee]{llava}
Haotian Liu, Chunyuan Li, Qingyang Wu, and Yong~Jae Lee.
\newblock Visual instruction tuning.
\newblock \emph{Advances in neural information processing systems},
  36:\penalty0 34892--34916, 2023{\natexlab{b}}.

\bibitem[Liu et~al.(2025{\natexlab{a}})Liu, Peng, Zhong, Yue, Lu, Yu, and
  Jia]{segzero}
Yuqi Liu, Bohao Peng, Zhisheng Zhong, Zihao Yue, Fanbin Lu, Bei Yu, and Jiaya
  Jia.
\newblock Seg-zero: Reasoning-chain guided segmentation via cognitive
  reinforcement.
\newblock \emph{arXiv preprint arXiv:2503.06520}, 2025{\natexlab{a}}.

\bibitem[Liu et~al.(2025{\natexlab{b}})Liu, Dong, Wang, Liu, Hu, Lu, and
  Rao]{ola}
Zuyan Liu, Yuhao Dong, Jiahui Wang, Ziwei Liu, Winston Hu, Jiwen Lu, and
  Yongming Rao.
\newblock Ola: Pushing the frontiers of omni-modal language model.
\newblock \emph{arXiv preprint arXiv:2502.04328}, 2025{\natexlab{b}}.

\bibitem[Muennighoff et~al.(2025)Muennighoff, Yang, Shi, Li, Fei-Fei,
  Hajishirzi, Zettlemoyer, Liang, Cand{\`e}s, and Hashimoto]{s1}
Niklas Muennighoff, Zitong Yang, Weijia Shi, Xiang~Lisa Li, Li~Fei-Fei,
  Hannaneh Hajishirzi, Luke Zettlemoyer, Percy Liang, Emmanuel Cand{\`e}s, and
  Tatsunori Hashimoto.
\newblock s1: Simple test-time scaling.
\newblock \emph{arXiv preprint arXiv:2501.19393}, 2025.

\bibitem[OpenAI(2023{\natexlab{a}})]{chatgpt}
OpenAI.
\newblock Chatgpt.
\newblock \url{https://openai.com/blog/chatgpt/}, 2023{\natexlab{a}}.

\bibitem[OpenAI(2023{\natexlab{b}})]{gpt4}
OpenAI.
\newblock Gpt-4 technical report.
\newblock \emph{arXiv:2303.08774}, 2023{\natexlab{b}}.

\bibitem[OpenAI(2025)]{gpt5}
OpenAI.
\newblock Gpt-5 system card, 2025.

\bibitem[OpenGVLab(2024)]{internomni}
OpenGVLab.
\newblock {InternOmni}: Extending internvl with audio modality.
\newblock \url{https://internvl.github.io/blog/2024-07-27-InternOmni/}, 2024.

\bibitem[Panayotov et~al.(2015)Panayotov, Chen, Povey, and
  Khudanpur]{libri-speech}
Vassil Panayotov, Guoguo Chen, Daniel Povey, and Sanjeev Khudanpur.
\newblock Librispeech: an asr corpus based on public domain audio books.
\newblock In \emph{2015 IEEE international conference on acoustics, speech and
  signal processing (ICASSP)}, pp.\  5206--5210. IEEE, 2015.

\bibitem[Radford et~al.(2021)Radford, Kim, Hallacy, Ramesh, Goh, Agarwal,
  Sastry, Askell, Mishkin, Clark, et~al.]{clip}
Alec Radford, Jong~Wook Kim, Chris Hallacy, Aditya Ramesh, Gabriel Goh,
  Sandhini Agarwal, Girish Sastry, Amanda Askell, Pamela Mishkin, Jack Clark,
  et~al.
\newblock Learning transferable visual models from natural language
  supervision.
\newblock In \emph{International conference on machine learning}, pp.\
  8748--8763. PmLR, 2021.

\bibitem[Radford et~al.(2022)Radford, Kim, Xu, Brockman, McLeavey, and
  Sutskever]{whisper}
Alec Radford, Jong~Wook Kim, Tao Xu, Greg Brockman, Christine McLeavey, and
  Ilya Sutskever.
\newblock Robust speech recognition via large-scale weak supervision, 2022.
\newblock URL \url{https://arxiv.org/abs/2212.04356}.

\bibitem[Shen et~al.(2018)Shen, Pang, Weiss, Schuster, Jaitly, Yang, Chen,
  Zhang, Wang, Skerrv-Ryan, et~al.]{shen2018natural}
Jonathan Shen, Ruoming Pang, Ron~J Weiss, Mike Schuster, Navdeep Jaitly,
  Zongheng Yang, Zhifeng Chen, Yu~Zhang, Yuxuan Wang, Rj~Skerrv-Ryan, et~al.
\newblock Natural tts synthesis by conditioning wavenet on mel spectrogram
  predictions.
\newblock In \emph{2018 IEEE international conference on acoustics, speech and
  signal processing (ICASSP)}, pp.\  4779--4783. IEEE, 2018.

\bibitem[Shi et~al.(2020)Shi, Bu, Xu, Zhang, and Li]{aishell3}
Yao Shi, Hui Bu, Xin Xu, Shaoji Zhang, and Ming Li.
\newblock Aishell-3: A multi-speaker mandarin tts corpus and the baselines.
\newblock \emph{arXiv preprint arXiv:2010.11567}, 2020.

\bibitem[Tang et~al.(2023)Tang, Yu, Sun, Chen, Tan, Li, Lu, Ma, and
  Zhang]{salmonn}
Changli Tang, Wenyi Yu, Guangzhi Sun, Xianzhao Chen, Tian Tan, Wei Li, Lu~Lu,
  Zejun Ma, and Chao Zhang.
\newblock Salmonn: Towards generic hearing abilities for large language models.
\newblock \emph{arXiv preprint arXiv:2310.13289}, 2023.

\bibitem[Team et~al.(2023)Team, Anil, Borgeaud, Alayrac, Yu, Soricut,
  Schalkwyk, Dai, Hauth, Millican, et~al.]{gemini}
Gemini Team, Rohan Anil, Sebastian Borgeaud, Jean-Baptiste Alayrac, Jiahui Yu,
  Radu Soricut, Johan Schalkwyk, Andrew~M Dai, Anja Hauth, Katie Millican,
  et~al.
\newblock Gemini: a family of highly capable multimodal models.
\newblock \emph{arXiv preprint arXiv:2312.11805}, 2023.

\bibitem[Team(2025)]{moss-ttsd}
OpenMOSS Team.
\newblock Text to spoken dialogue generation.
\newblock 2025.

\bibitem[Touvron et~al.(2023)Touvron, Lavril, Izacard, Martinet, Lachaux,
  Lacroix, Rozi{\`e}re, Goyal, Hambro, Azhar, Rodriguez, Joulin, Grave, and
  Lample]{llama}
Hugo Touvron, Thibaut Lavril, Gautier Izacard, Xavier Martinet, Marie-Anne
  Lachaux, Timoth{\'e}e Lacroix, Baptiste Rozi{\`e}re, Naman Goyal, Eric
  Hambro, Faisal Azhar, Aurelien Rodriguez, Armand Joulin, Edouard Grave, and
  Guillaume Lample.
\newblock Llama: Open and efficient foundation language models.
\newblock \emph{arXiv:2302.13971}, 2023.

\bibitem[Van Den~Oord et~al.(2016)Van Den~Oord, Dieleman, Zen, Simonyan,
  Vinyals, Graves, Kalchbrenner, Senior, Kavukcuoglu, et~al.]{van2016wavenet}
Aaron Van Den~Oord, Sander Dieleman, Heiga Zen, Karen Simonyan, Oriol Vinyals,
  Alex Graves, Nal Kalchbrenner, Andrew Senior, Koray Kavukcuoglu, et~al.
\newblock Wavenet: A generative model for raw audio.
\newblock \emph{arXiv preprint arXiv:1609.03499}, 12:\penalty0 1, 2016.

\bibitem[Wang et~al.(2024)Wang, Bai, Tan, Wang, Fan, Bai, Chen, Liu, Wang, Ge,
  et~al.]{qwen2vl}
Peng Wang, Shuai Bai, Sinan Tan, Shijie Wang, Zhihao Fan, Jinze Bai, Keqin
  Chen, Xuejing Liu, Jialin Wang, Wenbin Ge, et~al.
\newblock Qwen2-vl: Enhancing vision-language model's perception of the world
  at any resolution.
\newblock \emph{arXiv preprint arXiv:2409.12191}, 2024.

\bibitem[Wu et~al.(2023)Wu, Waheed, Zhang, Abdul-Mageed, and Aji]{lamini-lm}
Minghao Wu, Abdul Waheed, Chiyu Zhang, Muhammad Abdul-Mageed, and Alham~Fikri
  Aji.
\newblock Lamini-lm: A diverse herd of distilled models from large-scale
  instructions.
\newblock \emph{CoRR}, abs/2304.14402, 2023.
\newblock URL \url{https://arxiv.org/abs/2304.14402}.

\bibitem[Xie \& Wu(2024{\natexlab{a}})Xie and Wu]{mini-omni}
Zhifei Xie and Changqiao Wu.
\newblock Mini-omni: Language models can hear, talk while thinking in
  streaming.
\newblock \emph{arXiv preprint arXiv:2408.16725}, 2024{\natexlab{a}}.

\bibitem[Xie \& Wu(2024{\natexlab{b}})Xie and Wu]{mini-omni2}
Zhifei Xie and Changqiao Wu.
\newblock Mini-omni2: Towards open-source gpt-4o with vision, speech and duplex
  capabilities.
\newblock \emph{arXiv preprint arXiv:2410.11190}, 2024{\natexlab{b}}.

\bibitem[Xu et~al.(2025)Xu, Guo, He, Hu, He, Bai, Chen, Wang, Fan, Dang,
  et~al.]{qwen25omni}
Jin Xu, Zhifang Guo, Jinzheng He, Hangrui Hu, Ting He, Shuai Bai, Keqin Chen,
  Jialin Wang, Yang Fan, Kai Dang, et~al.
\newblock Qwen2. 5-omni technical report.
\newblock \emph{arXiv preprint arXiv:2503.20215}, 2025.

\bibitem[Yang et~al.(2025{\natexlab{a}})Yang, Li, Yang, Zhang, Hui, Zheng, Yu,
  Gao, Huang, Lv, et~al.]{qwen3}
An~Yang, Anfeng Li, Baosong Yang, Beichen Zhang, Binyuan Hui, Bo~Zheng, Bowen
  Yu, Chang Gao, Chengen Huang, Chenxu Lv, et~al.
\newblock Qwen3 technical report.
\newblock \emph{arXiv preprint arXiv:2505.09388}, 2025{\natexlab{a}}.

\bibitem[Yang et~al.(2025{\natexlab{b}})Yang, Yang, Zhang, Hui, Zheng, Yu, Li,
  Liu, Huang, et~al.]{qwen2.5}
An~Yang, Baosong Yang, Beichen Zhang, Binyuan Hui, Bo~Zheng, Bowen Yu,
  Chengyuan Li, Dayiheng Liu, Fei Huang, et~al.
\newblock Qwen2.5 technical report.
\newblock \emph{arXiv preprint arXiv:2502.13923}, 2025{\natexlab{b}}.

\bibitem[Yang et~al.(2024)Yang, Xu, Liu, Chu, Jiang, Zhou, Leng, Lv, Zhao,
  Zhou, et~al.]{airbench}
Qian Yang, Jin Xu, Wenrui Liu, Yunfei Chu, Ziyue Jiang, Xiaohuan Zhou, Yichong
  Leng, Yuanjun Lv, Zhou Zhao, Chang Zhou, et~al.
\newblock Air-bench: Benchmarking large audio-language models via generative
  comprehension.
\newblock \emph{arXiv preprint arXiv:2402.07729}, 2024.

\bibitem[Zhang et~al.(2023)Zhang, Li, Zhang, Zhan, Wang, Zhou, and
  Qiu]{speechgpt}
Dong Zhang, Shimin Li, Xin Zhang, Jun Zhan, Pengyu Wang, Yaqian Zhou, and
  Xipeng Qiu.
\newblock Speechgpt: Empowering large language models with intrinsic
  cross-modal conversational abilities.
\newblock \emph{arXiv preprint arXiv:2305.11000}, 2023.

\bibitem[Zhang et~al.(2024)Zhang, Dong, Cao, Zang, Qian, Wei, Chen, Li, Niu,
  Ding, et~al.]{internlm}
Pan Zhang, Xiaoyi Dong, Yuhang Cao, Yuhang Zang, Rui Qian, Xilin Wei, Lin Chen,
  Yifei Li, Junbo Niu, Shuangrui Ding, et~al.
\newblock Internlm-xcomposer2. 5-omnilive: A comprehensive multimodal system
  for long-term streaming video and audio interactions.
\newblock \emph{arXiv preprint arXiv:2412.09596}, 2024.

\bibitem[Zhong et~al.(2024)Zhong, Wang, Liu, Yang, Tang, Zhang, Li, Qu, Li,
  Chen, et~al.]{lyra}
Zhisheng Zhong, Chengyao Wang, Yuqi Liu, Senqiao Yang, Longxiang Tang, Yuechen
  Zhang, Jingyao Li, Tianyuan Qu, Yanwei Li, Yukang Chen, et~al.
\newblock Lyra: An efficient and speech-centric framework for omni-cognition.
\newblock \emph{arXiv preprint arXiv:2412.09501}, 2024.

\bibitem[Zhu et~al.(2023)Zhu, Chen, Shen, Li, and Elhoseiny]{minigpt4}
Deyao Zhu, Jun Chen, Xiaoqian Shen, Xiang Li, and Mohamed Elhoseiny.
\newblock Minigpt-4: Enhancing vision-language understanding with advanced
  large language models.
\newblock \emph{arXiv preprint arXiv:2304.10592}, 2023.

\end{thebibliography}
\bibliographystyle{iclr2026_conference}

\clearpage
\appendix
\section{Appendix}

\subsection{Data Format}
\begin{figure}[h]
\begin{center}
\includegraphics[width=\textwidth]{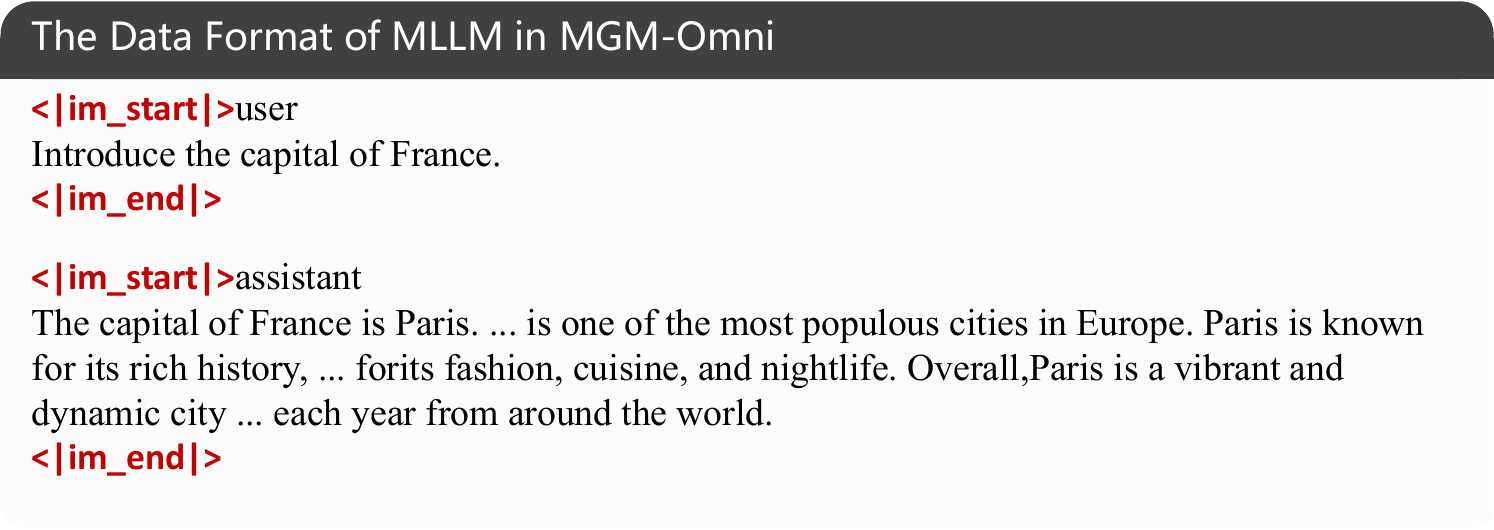}
\hspace{5mm}
\includegraphics[width=\textwidth]{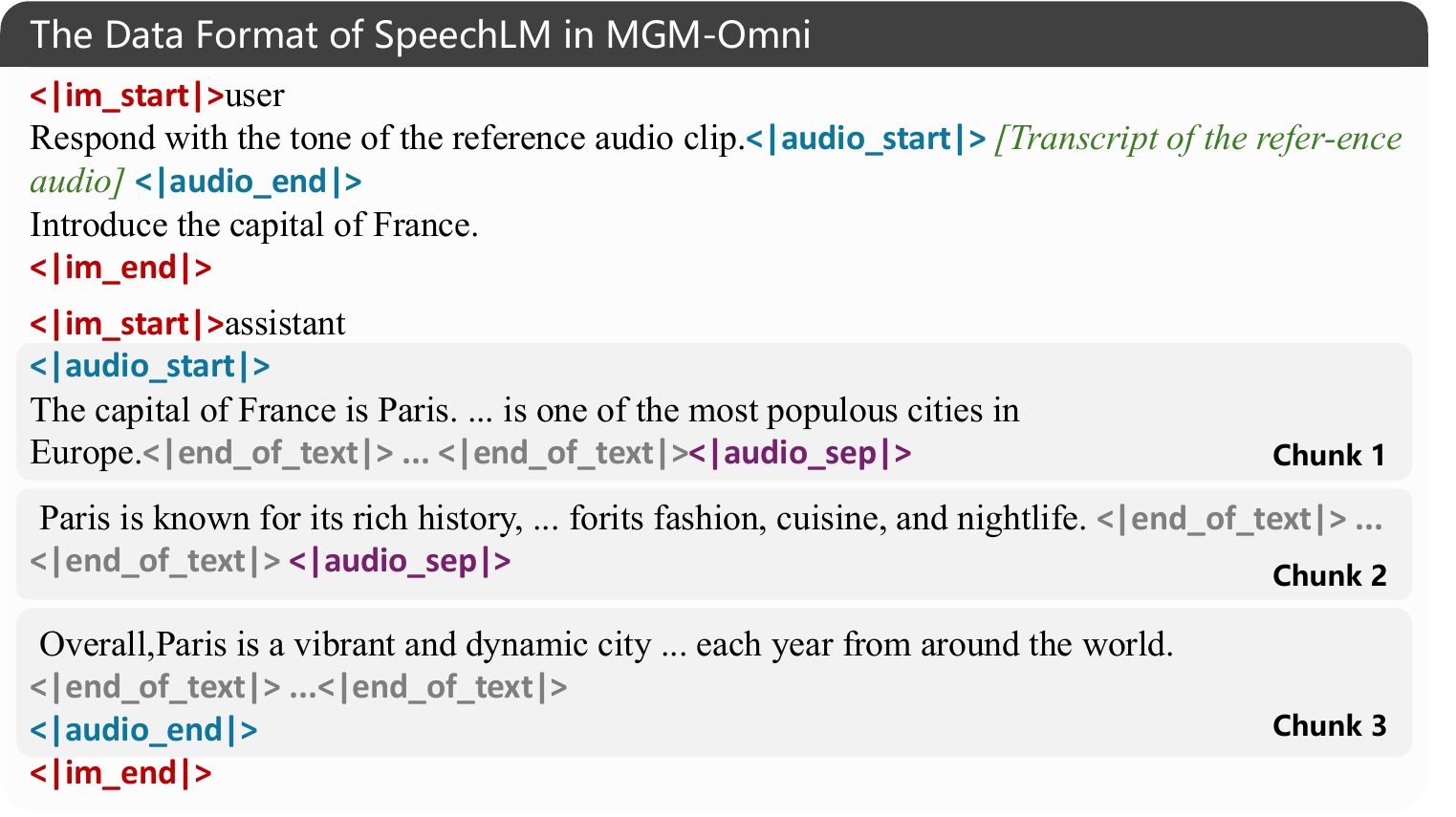}
\caption{The data format of MLLM (top) and SpeechLM (button) in MGM-Omni.}
\label{data_format}
\end{center}
\end{figure}

The data format for MLLM and SpeechLM with the same instruction is illustrated in Figure~\ref{data_format}. SpeechLM use chunk-based decoding to generate long-form speech.

\subsection{Long-TTS-Eval Benchmark}
\label{sec:long-tts-eval}

In this section, we provide a detailed introduction to the data composition and evaluation protocol of the Long-TTS-Eval benchmark we constructed.

\subsubsection{Data Composition}
Long-TTS-Eval focuses on assessing TTS systems' capabilities in long-form speech generation and complex case handling. 

For long TTS evaluation, we collected six types of text: literature, news, knowledge, speeches, reviews, and academic papers, comprising 341 Chinese samples and 353 English samples. The data were sourced from news outlets, Wikipedia, YouTube video transcripts, and arXiv papers.
We use the Qwen3 tokenizer~\citep{qwen3} to calculate the token length. As illustrated in Tabel~\ref{tab:long-tts-info}, the maximum length is 1899 tokens in Chinese and 3277 tokens in English, and the average length is 689.57 tokens in Chinese and 1019.0 tokens in English. As a single-point timing estimate, 1899 Chinese tokens correspond to about 10 minutes of speech (assuming 200 characters per minute and 1 token per character), and 3277 English tokens correspond to about 12 minutes (assuming 215 words per minute and 1.3-1.5 tokens per word).

For complex case handling, we collected five types of text: web URLs, emails, math formulas, phone numbers, and large numbers, comprising 265 Chinese samples and 260 English samples. The detailed information is illustrated in Tabel~\ref{tab:hard-tts-info}. Mathematical formulas were sourced from the reasoning process and solution from S1 Long-CoT Instruct dataset~\cite{s1}, while the other categories were generated by Gemini 2.5 Pro~\citep{gemini2.5}.

\subsubsection{Evaluation Pipeline}
We follow Seed-TTS-Eval~\citep{seedtts} to build our evaluation pipeline. We use Whisper-large-v3~\citep{whisper} and Paraformer-zh~\citep{funasr} as the automatic speech recognition (ASR) engines for English and Chinese, respectively. Since both models accept only short audio, we segment each generated waveform into 28-second chunks, transcribe each chunk independently, and then concatenate the transcripts to obtain the final transcription. We then compute word error rate (WER) for English and character error rate (CER) for Chinese.

\subsubsection{Evaluation with Normalized Text}
Conventional TTS benchmarks often transcribe generated speech with an ASR model and then compare the transcript to the ground-truth text to calculate the error rate. This approach has a key flaw: for expressions with multiple valid readings, ASR outputs can legitimately differ from the written form. For example, "5\%" spoken by TTS may be transcribed as "five percent." It differs from the ground truth, but it is still correct.

To address this issue, for each sample with ground-turth $G$, we prompt GPT-5~\citep{gpt5} to generate a normalized ground-turth $N$ that reflects a natural spoken version. We then synthesize speech, obtain the ASR transcript $T$, and compute two word error rates, between $T$ and $G$, and between $T$ and $N$. The final per-sample error is the smaller of the two:
\begin{equation}
    WER_{sample} = min(WER(T, G), WER(T, N))
\end{equation}
This method lowers the risk of falsely flagging correct TTS, thereby enhancing the reliability of the reported error rates.

\begin{table}[t]
\centering
\resizebox{\columnwidth}{!}{%
\begin{tabular}{ccccccc}
\toprule
Category & Samples (ZH) & Avg Length (ZH) & Max Length (ZH) & Samples (EN) & Avg Length (EN) & Max Length (EN) \\ \midrule
Literature & 41  & 998.8  & 1644 & 56  & 985.5  & 1344 \\
News       & 60  & 585.4  & 1159 & 60  & 915.4  & 1781 \\
Knowledge  & 60  & 764.0  & 1279 & 59  & 1130.7 & 3245 \\
Talk       & 60  & 619.8  & 1885 & 59  & 952.4  & 2745 \\
Comment    & 60  & 513.8  & 1537 & 59  & 844.6  & 2096 \\
Paper      & 60  & 753.5  & 1899 & 60  & 1281.2 & 3277 \\ \midrule
Total      & 341 & 689.6  & 1899 & 353 & 1019.0 & 3277 \\ \bottomrule
\end{tabular}%
}
\caption{The composition and average length of our Long-TTS-Eval benchmark.}
\label{tab:long-tts-info}
\end{table}

\begin{table}[t]
\centering
\resizebox{\columnwidth}{!}{%
\begin{tabular}{ccccccc}
\toprule
Category & Samples (ZH) & Avg Length (ZH) & Max Length (ZH) & Samples (EN) & Avg Length (EN) & Max Length (EN) \\ \midrule
URLs    & 57  & 96.8  & 180 & 45  & 102.5 & 166  \\
Emails  & 45  & 63.6  & 97  & 44  & 85.5  & 136  \\
Phone   & 30  & 92.0  & 160 & 30  & 117.2 & 199  \\
Number  & 33  & 83.7  & 159 & 30  & 94.0  & 136  \\
Math    & 100 & 606.8 & 955 & 100 & 605.8 & 1009 \\ \midrule
Total   & 265 & 281.4 & 955 & 260 & 293.9 & 1009 \\ \bottomrule
\end{tabular}%
}
\caption{The composition and average length of the hard set in our Long-TTS-Eval benchmark.}
\label{tab:hard-tts-info}
\end{table}

\subsection{Quantitative Results}

\subsubsection{Long Audio Understanding}

\begin{figure}[]
\begin{center}
\includegraphics[width=1.0\textwidth]{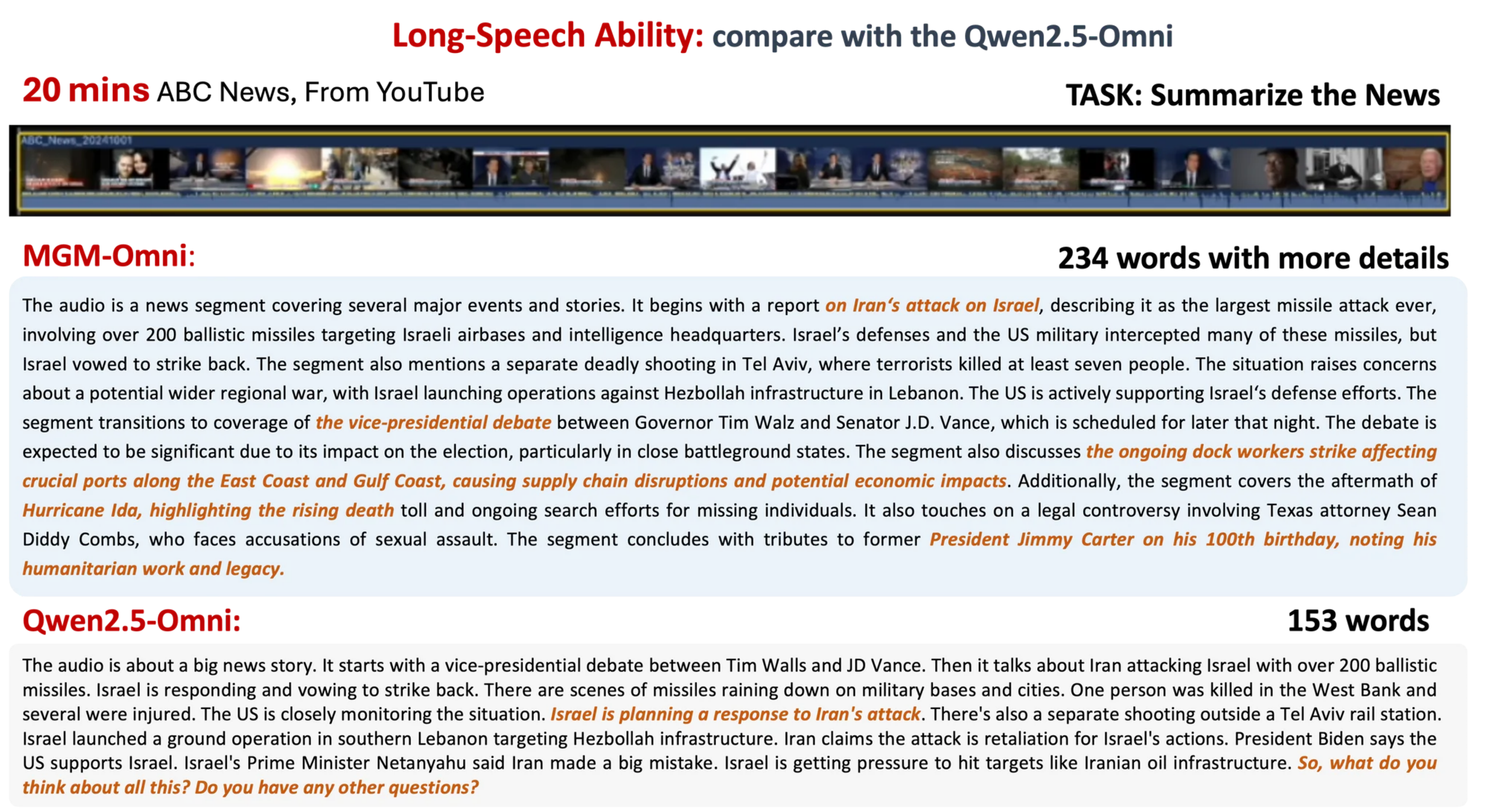}
\includegraphics[width=1.0\textwidth]{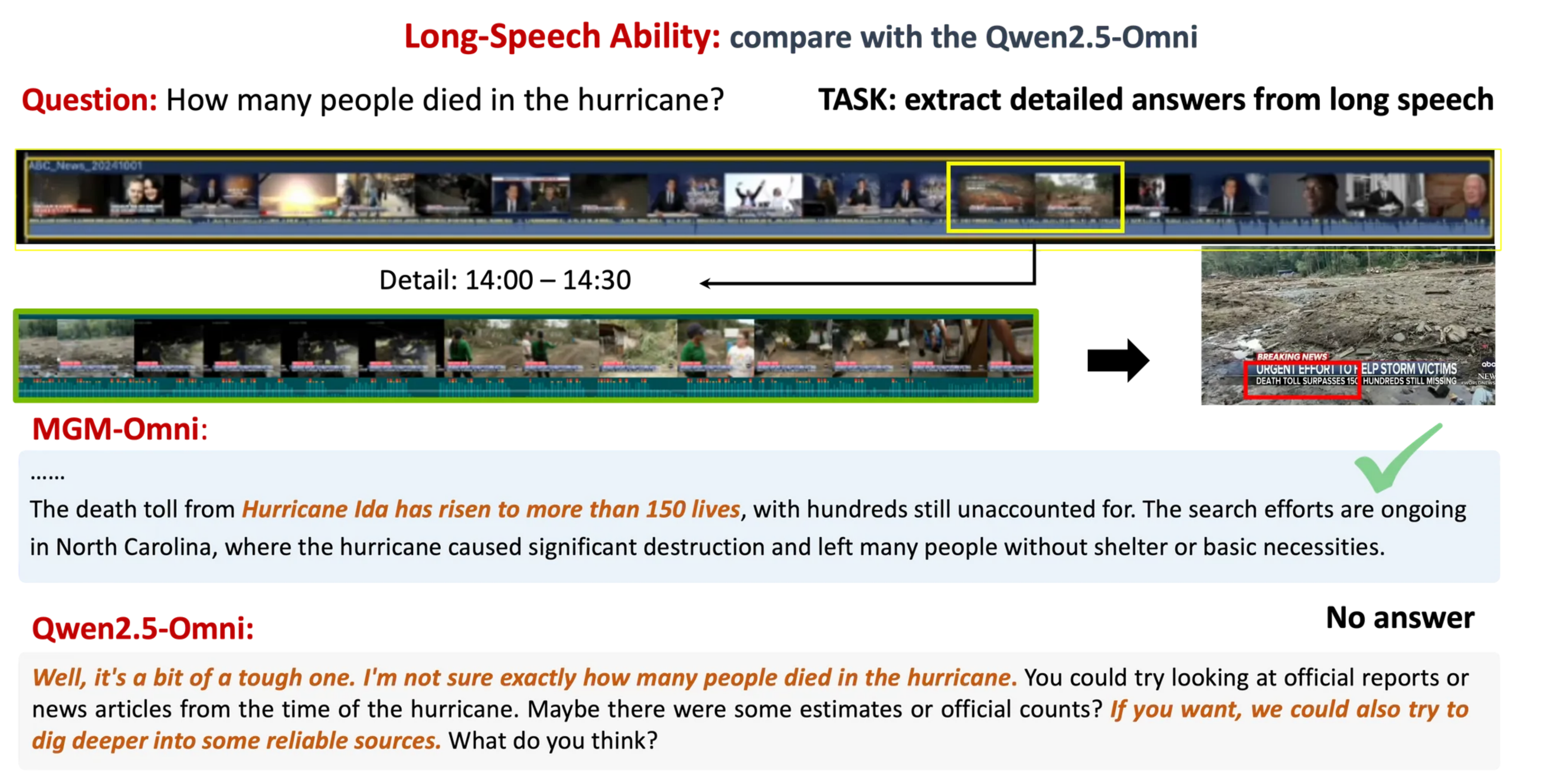}
\caption{MGM-Omni is capable of understanding long-form audio.}
\label{fig:long-asr}
\end{center}
\end{figure}

To verify MGM-Omni's effectiveness in long audio understanding, we conducted a more in-depth evaluation. We illustrate the quantitative result in Figure~\ref{fig:long-asr}. For long audio summarization, MGM-Omni provides more complete and detailed responses compared with Qwen2.5-Omni~\citep{qwen25omni}. For fine-grained understanding, MGM-Omni accurately extracts information from long audio inputs, while Qwen2.5-Omni refuses to respond.

\subsubsection{Long Speech Generation}

\begin{figure}[]
\begin{center}
\includegraphics[width=1.0\textwidth]{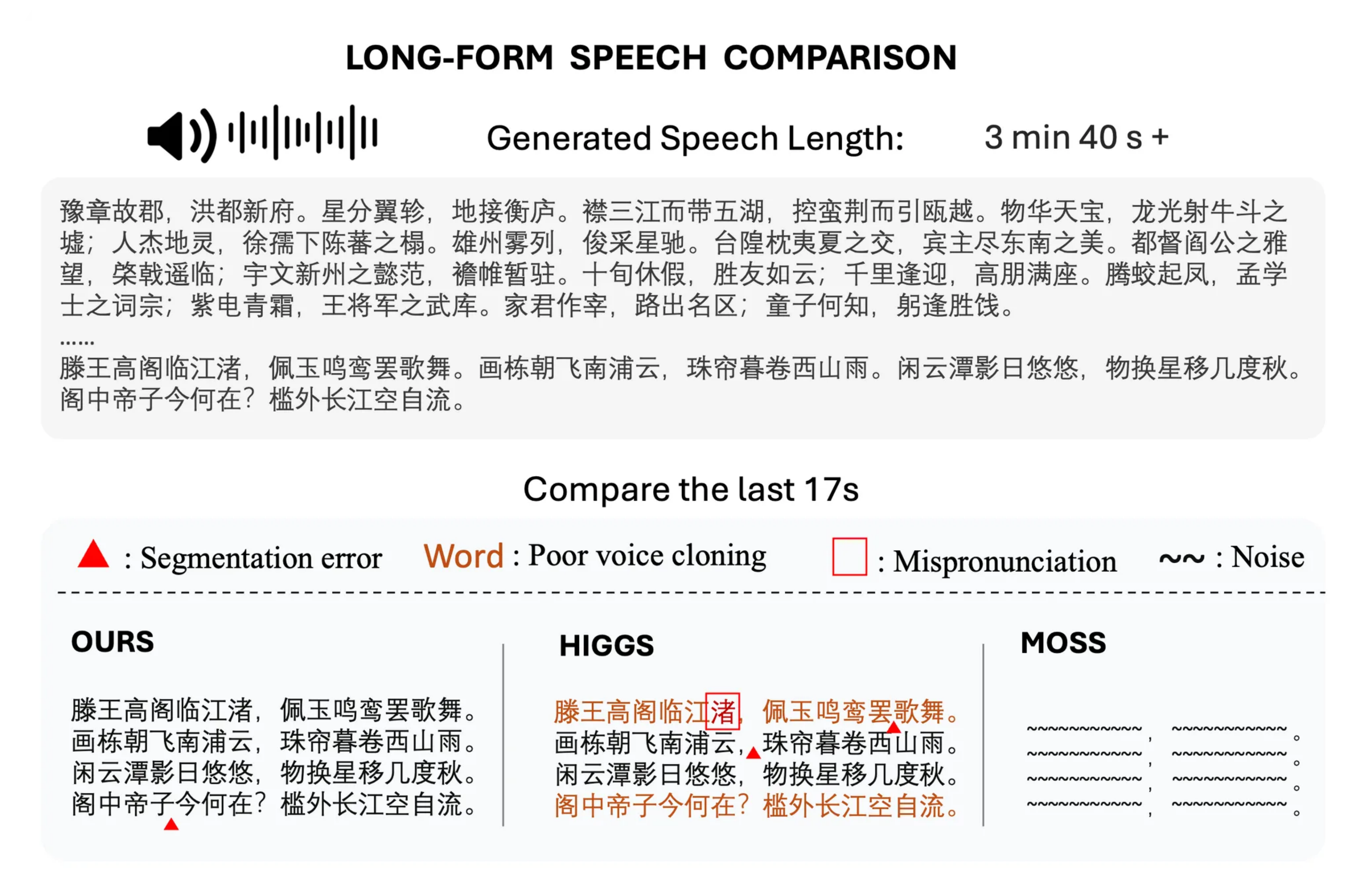}
\includegraphics[width=1.0\textwidth]{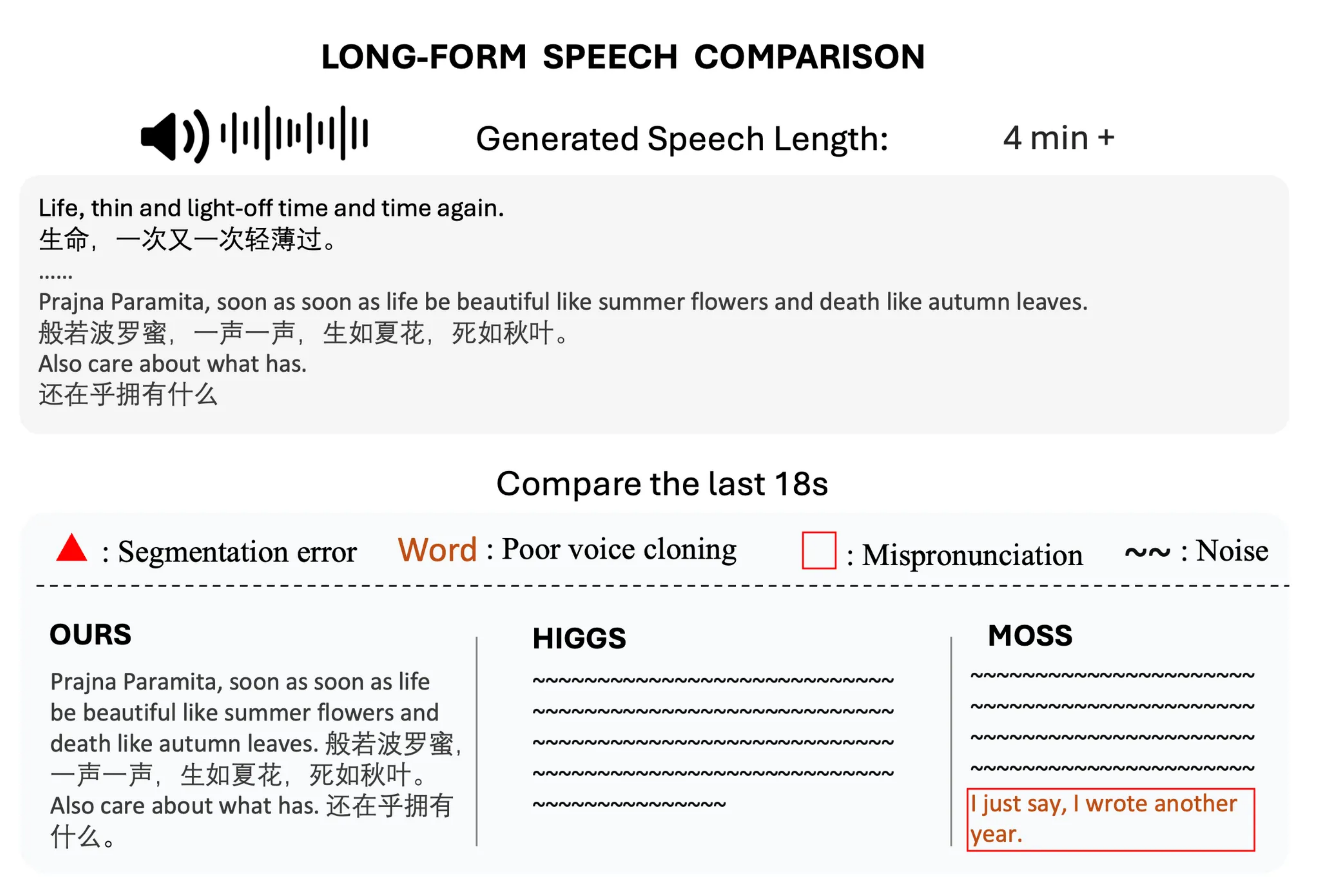}
\caption{MGM-Omni is capable of correctly generating long-form speech.}
\label{fig:long-tts}
\end{center}
\end{figure}

We compare MGM-Omni with concurrent long TTS systems, MOSS-TTSD-v0.5~\citep{moss-ttsd} and Higgs-Audio-v2~\citep{higgs-audio} to evaluate the long-form speech generation capability. Specifically, we evaluate two challenging pieces: the renowned Chinese long prose poem "Preface to the Pavilion of Prince Teng" (Tengwang Ge Xu) and Tagore's famous poem "Stray Birds" excerpt "Life is as ephemeral as summer flowers" featuring mixed Chinese-English code switching. As depicted in Figure~\ref{fig:long-tts}, MGM-Omni produces accurate speech with appropriate pausing, while competing methods exhibit pronounced errors in the latter portions of the audio, including audible noise.

\subsection{Use of LLMs}
In this study, we utilize large language models (LLMs), specifically GPT-5~\citep{gpt5}, to enhance the quality of our writing by correcting grammatical errors, improving sentence structure, and refining overall clarity. All ideas, methodologies, experimental designs, and results are entirely the original work of the authors, with LLMs serving solely as tools for language enhancement.

\clearpage

\end{document}